\documentclass[letter,twocolumn]{jpsj3}
\usepackage{txfonts}
\usepackage{braket}
\usepackage{dcolumn}
\usepackage{bm}
\usepackage{color}
\usepackage[dvipdfmx]{graphicx}

\title{Study of Staggered Magnetization in the Spin-$S$ Square-Lattice Heisenberg Model Using Spiral Boundary Conditions}

\author{Masahiro Kadosawa$^1$\thanks{afna1728@chiba-u.jp}, Masaaki Nakamura$^2$, Yukinori Ohta$^1$, and Satoshi Nishimoto$^{3,4}$}
\inst{$^1$Department of Physics, Chiba University, Chiba 263-8522, Japan\\
$^2$Department of Physics, Ehime University, Ehime 790-8577, Japan\\
$^3$Department of Physics, Technical University Dresden, 01069 Dresden, Germany \\ 
$^4$Institute for Theoretical Solid State Physics, IFW Dresden, 01171 Dresden, Germany
} 

\abst{
We propose an efficient numerical method to obtain local order
parameter in two-dimensional systems using spiral boundary
conditions. As a benchmark, we first estimate the magnitude of
staggered magnetization for the $S=1/2$ XXZ Heisenberg model
on a square lattice in the whole range of the XXZ anisotropy
by density-matrix renormalization group technique.
The validity of our method is confirmed by comparing our results
with the previous analytical and numerical studies. Then, as
further demonstration, we apply our method to obtain the
staggered magnetization for the higher-spin cases from $S=1$ to
$S=6$. The accuracy of the obtained results is validated using
the series expansion and the spin-wave theory.
}

\begin{document}
\maketitle

The Heisenberg model~\cite{Heisenberg1928} is used to study the magnetic properties of various materials. The geometry of spin network as well as the exchange anisotropy ($\Delta$) are
key factors to determine the magnetic ground state. Such
anisotropy often exists in real materials due to the effects of
crystal fields and so on~\cite{Lines1963,Achiwa1969}. While
easy-axis anisotropy typically reduces quantum fluctuations,
easy-plane one brings a two-dimensional (2D) system
to the BKT universality class~\cite{Kosterlitz1973}.
Interestingly, recent studies reported a possible tuning of
$\Delta$ by magnetic field in
[Cu(pz)$_2$(2-HOpy)$_2$](PF$_6$)$_2$~\cite{Opherden2022} and a
switching between $\Delta>1$ and $\Delta<1$ by coligands in
cobalt complexes~\cite{Wu2021}. In particular, easy-plane
magnets provide an exciting platform for studying topological
excitations referred to as vortices~\cite{Sonin2010}.

Another important factor is the magnitude of spin ($S$).
One may think that a quantum system just approaches the
classical limit with increasing $S$. However, the actual physics
is not so simple because of the presence of specific features
like the Haldane state. In fact, even nowadays, fascinating
experimental measurements for high-$S$ materials have been
successively reported: For example,
square magnets Ba$_2$FeSi$_2$O$_7$
($S=2$)~\cite{Jang2021,Do2022,Lee2022},
NaMnSbO$_4$ ($S=5/2$)~\cite{Vasilchikova2020};
triangular magnets
PbMnTeO$_6$ ($S=3/2$)~\cite{Kuchugura2019},
AgCrSe$_2$ ($S=3/2$)~\cite{Baenitz2021},
$\alpha$-CrOO(H,D) ($S=3/2$)~\cite{Liu2021}
Ba$_8$MnNb$_6$O$_{24}$ ($S=5/2$)~\cite{Rawl2019},
Ag$_2$FeO$_2$ ($S=5/2$)~\cite{Yoshida2020};
honeycomb magnet CoPS$_3$ ($S=3/2$)~\cite{Kim2020};
kagome magnets Cs$_2$(K,Na)Cr$_3$F$_{12}$ ($S=3/2$)~\cite{Goto2018},
Li$_9$Fe$_3$(P$_2$O$_7$)$_3$(PO$_4$)$_2$ ($S=5/2$)~\cite{Kermarrec2021},
PbFe$_3$(PO$_4$)(SO$_4$)(OH)$_6$ ($S=5/2$)~\cite{Ferrenti2022}.
Furthermore, the exploration of spin-liquid ground state is
recently heating up with the appearance of high-$S$ Kitaev
materials~\cite{Sano2018,Lee2020,Xu2020}.

Under these circumstances, numerical techniques to systematically
study the high-$S$ Heisenberg systems in a wide range of
$\Delta$ are increasingly required. However, the computational
cost would be significant for such cases, so that it is often
hard to obtain quantities accurately in the thermodynamic
limit. Thus, in this paper, we propose an efficient
method to obtain order parameter for 2D systems with
spiral boundary conditions (SBC). Using SBC, a 2D system can be
exactly projected onto a one-dimensional (1D) periodic chain
with translation symmetry. This enables us to perform optimal
DMRG calculations and simple finite-size scaling.
As a benchmark, we estimate the magnitude of staggered
magnetization for $S=1/2$ XXZ square-lattice Heisenberg model
by density-matrix renormalization group (DMRG).
To confirm the validity of our method, the results are
compared to the previous studies by DMRG~\cite{White2007}, 
quantum Monte Carlo (QMC)~\cite{Sandvik1999,Sandvik2010},
spin wave theory
(SWT)~\cite{Hamer1992,Igarashi1992,Canali1993}, series expansion
(SE) ~\cite{Davis1960,Huse1988,Parrinello1974,Singh1989,Singh1990,Weihong1991-1,Weihong1991-2,Hamer1991}, and coupled-cluster
method (CCM)~\cite{BISHOP2017}.
We then demonstrate that this method can achieve a high
performance level even for high-$S$ cases.

The Hamiltonian of the spin-$S$ XXZ model on a square lattice
is written as
$\mathcal{H} = \sum_{\braket{ij}}(S^{x}_{i}S^{x}_{j} + S^{y}_{i}S^{y}_{j} + \Delta S^{z}_{i}S^{z}_{j})$,
where $S^\gamma_i$ $(\gamma = x,y,z)$ are the spin-$S$ operators,
$\Delta$ is the anisotropy parameter, and the sum $\braket{ij}$
runs over all nearest-neighbor pairs. It is known that this
model exhibits long-range order for any $S$ and
$\Delta$~\cite{Kubo1988}. There are three phases depending on
$\Delta$~\cite{Kubo1988,Viswanath1994,Yunoki2002,Braiorr-Orrs2019}:
(i) For $\Delta > 1$ easy-axis N\'eel phase with
antiferromagnetic (AFM) spin alignment along the $z$-direction,
(ii) for $-1 < \Delta < 1$ easy-plane N\'eel (XY) phase with
AFM spin alignment along some arbitrary direction in the
$xy$-plane, and (iii) for $\Delta < -1$ ferromagnetic (FM)
phase with fully-polarized spins along the $z$-direction.
The phase transitions at $\Delta=\pm1$ are both first order.
For $\Delta=-1$ this model can be exactly solved: The ground
state is two-fold degenerate with energy $E_0=-2NS^2$. One of them is expressed as
$|\Psi_0({\rm XY})\rangle=\sum_m \lambda_m |\psi_m \rangle$,
where $|\psi_m \rangle$ are bases restricted to
$S^z_{\rm tot}=\sum_{i=1}^L \langle S^z_i \rangle=0$
subspace, $m$ is summed over all possible combinations of the
spin configurations, and $\lambda_m$ are determined for
each $S$ (see Supplementary data~\cite{SM}). The magnitude of staggered
magnetization is $S$ with the direction parallel to the
$xy$-plane. The other is a FM state
$|\Psi_0({\rm FM})\rangle=\frac{1}{\sqrt{2}}(|\Uparrow \rangle+|\Downarrow \rangle)$, where $|\Uparrow \rangle$ and
$|\Downarrow \rangle$ denote fully-polarized states toward
$z$ and $-z$ directions, respectively. These states 
$|\Psi_0({\rm XY})\rangle$ and $|\Psi_0({\rm FM})\rangle$ are orthogonal. Also, the system is simplified to an Ising model
at $\Delta\to\infty$ (Ising limit).

\begin{figure}[tb]
	\centering
	\includegraphics[width=0.9\linewidth]{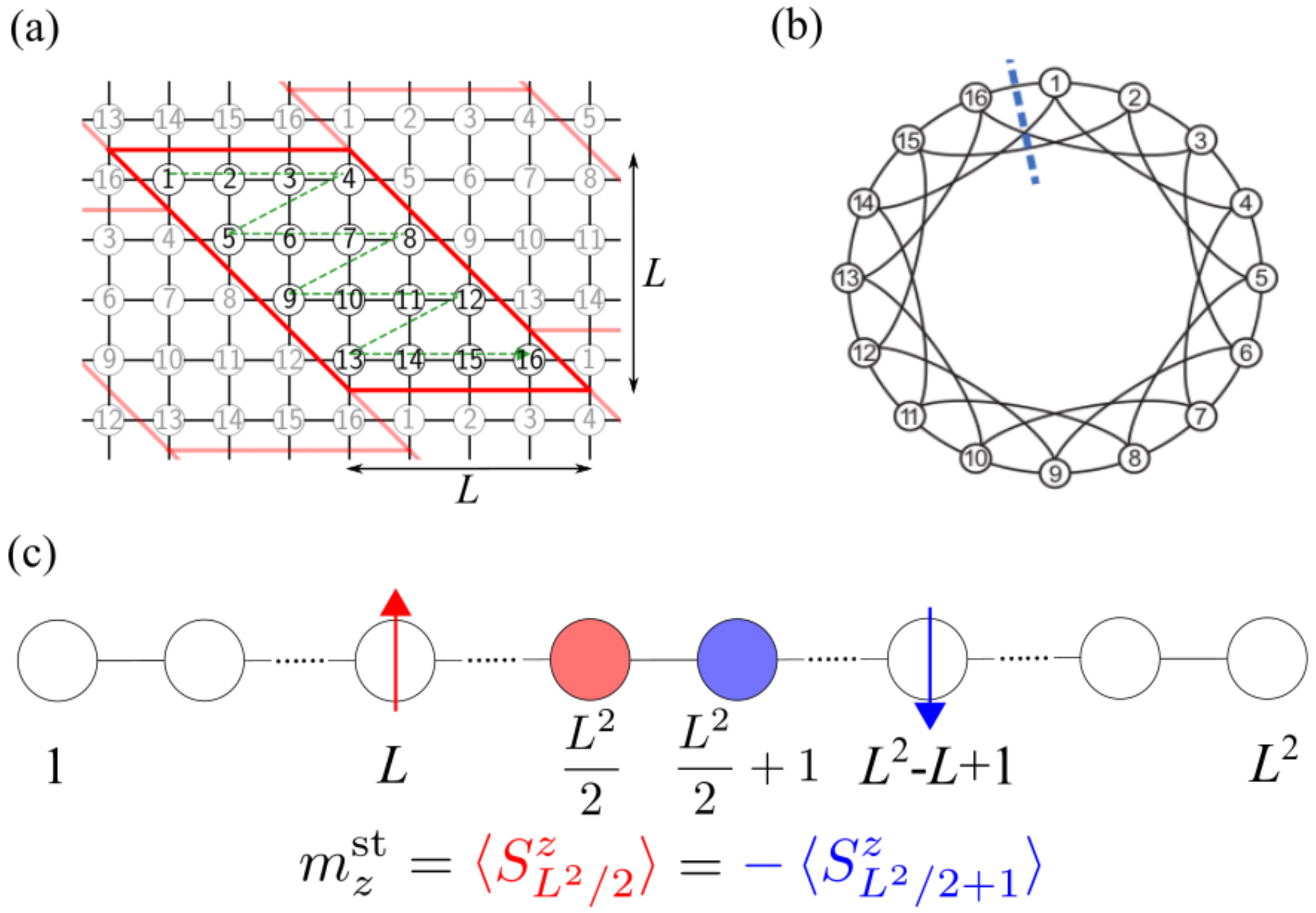}
    \caption{(Color online) (a) 2D square lattice with $L = 4$, where the
	region framed by red line is the original cluster.
	(b) 1D representation of (a) by numbering sites
	along green line. An open chain is created by cutting $L$
	bonds between two sites (dotted line).
	(c) Schematic picture of 1D open chain used for the DMRG
	calculations of $m_\alpha^{\mathrm{st}}$. The arrows denote
	pinned spins and $m_z^{\mathrm{st}}$ is measured at the
	filled sites.
	}
	\label{fig:SBC}
\end{figure}

First, we explain our method for estimating the magnitude
of magnetization. In general, it is difficult to estimate
an order parameter for 2D system with DMRG because not only
the implementation itself is challenging but also finite-size
scaling analysis is not straightforward. We manage to resolve
these issues by mapping 2D cluster onto 1D translation-symmetric
chain using SBC. Here, the original 2D square lattice with
$L\times L$ sites is mapped onto a 1D chain with nearest- and
($L-1$)-th neighbor bonds~\cite{Nakamura2021,Kadosawa2022}.
An example of this mapping scheme for a $4\times 4$ cluster
is illustrated in Fig.~\ref{fig:SBC}(a,b). Thus, the
Hamiltonian is translated to
$\mathcal{H} = \sum_{i=1}^{L^2}\left(S^{x}_{i}S^{x}_{i+1} + S^{y}_{i}S^{y}_{i+1} + \Delta S^{z}_{i}S^{z}_{i+1}\right)
	+ \sum_{i=1}^{L^2}\left(S^{x}_{i}S^{x}_{i+(L-1)} + S^{y}_{i}S^{y}_{i+(L-1)} + \Delta S^{z}_{i}S^{z}_{i+(L-1)}\right)$.
As a result, quantum entanglement is uniformly distributed
over the projected 1D chain due to the translation symmetry.
It is also important that the distance of the longest bonds
is minimized to be $L-1$. These conditions enable us to
optimally perform DMRG calculations. In addition, since the
lattice site is indexed by a single coordinate $i$ instead of
two coordinates $(i,j)$ in 2D cluster, just one finite-size
scaling analysis is required to obtain physical quantity in
the thermodynamic limit.

For an N\'eel state an order parameter is the magnitude
of staggered magnetization. Since the N\'eel order with
$\bm{k}=(\pi,\pi)$ in the original 2D representation
is expressed as that with $k=\pi$ along the projected
1D chain, the order parameter may be defined by 
retaining half-amplitude of the Friedel oscillation of
$\braket{S^\alpha_i}$ ($\alpha=x$, $y$, or $z$) starting
from the system edges if we use an open chain. Such an open
chain can be created by cutting $L$ bonds between two
sites of the projected 1D periodic chain [see Fig.~\ref{fig:SBC}(b)]. In practice, we measure the
local moments of central spins
$\braket{S^{z}_{L^2/2}}$, $\braket{S^{z}_{L^2/2+1}}$
($\braket{S^{x}_{L^2/2}}$, $\braket{S^{x}_{L^2/2+1}}$)
with pinning two spins near the system edges like 
$\braket{S^z_L}=1/2$, $\braket{S^z_{L^2-L+1}}=-1/2$
($\braket{S^x_L}=1/2$, $\braket{S^x_{L^2-L+1}}=-1/2$)
in the easy-axis (easy-plane) N\'eel phase. Thus,
the order parameters for the easy-axis and easy-plane
N\'eel states are defined as
$m_{z}^{\mathrm{st}}=|\braket{S^{z}_{L^2/2}}-\braket{S^{z}_{L^2/2+1}}|/2$ and
$m_{x}^{\mathrm{st}}=|\braket{S^{x}_{L^2/2}}-\braket{S^{x}_{L^2/2+1}}|/2$, respectively [see Fig.~\ref{fig:SBC}(c)].
Typically, such a pinning may be naively placed at system edges,
i.e., at $i=1$, $i=L^2$. However, since outer $L-1$ sites for
both edges lose the original bonds, the pinnings are placed
at the inner sites $i=L$, $i=L^2-L+1$ to avoid an
underestimate of order parameter (also see below).

We use DMRG method~\cite{White1992} to calculate the magnitude
of staggered magnetization. We study open chains with length
up to $N = L^2 = 196$ sites. We keep up to $\chi=8000$
density-matrix eigenstates, and the discarded weight is on
the order of $10^{-5}$ at most. The calculated values are
extrapolated to $\chi\to\infty$ if necessary. 
More detailed data on the accuracy of our DMRG
calculations are given in the Supplemental Material~\cite{SM}. For comparison,
the performances using various boundary conditions are also shown.

\begin{figure}[tb]
	\centering
	\includegraphics[width=0.9\linewidth]{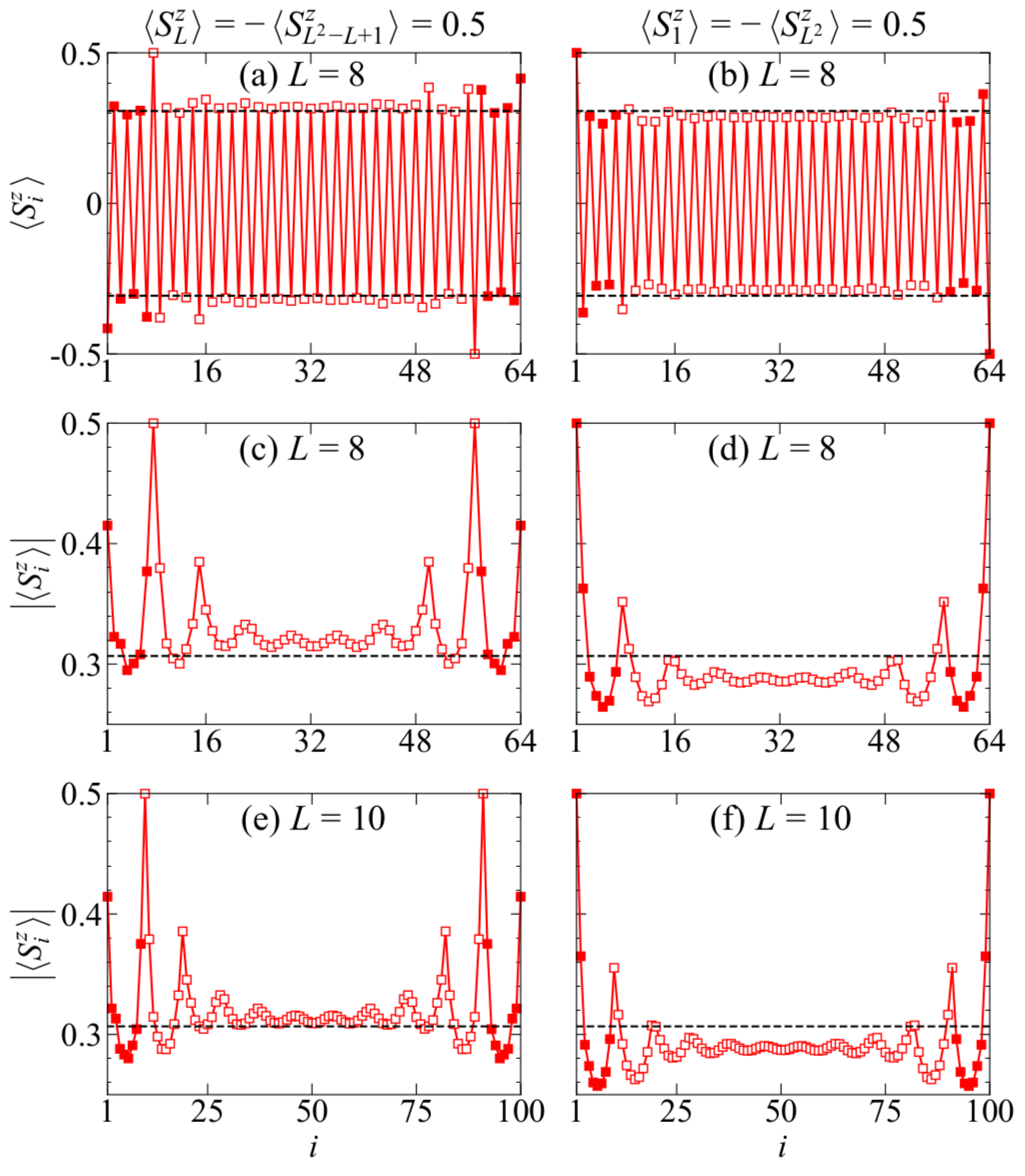}
    \caption{(Color online) (a) Local spin moment $\braket{S^z_i}$ and (c,e)
	$|\braket{S^z_i}|$ with pinning at $i=L$, $i=L^2-L+1$.
	(b,d,f) The same quantities with pinning at $i=1$, $i=L^2$.
	Dashed lines ($m_{z}^{\mathrm{st}}=0.3067$~\cite{White2007})
	are guide for the convergence. The solid squares correspond
	to sites which lose the original bonds (see text).
	}
	\label{fig:Szi}
\end{figure}

As an illustration, we plot the Friedel oscillation of
$\braket{S^{z}_{i}}$ for $S=1/2$ and $\Delta=1$
in Fig.~\ref{fig:Szi}(a), where spins at $i=L$, $i=L^2-L+1$
are pinned. A staggered oscillation with $k=\pi$ is obviously
seen. Also, the oscillation of $\braket{S^{z}_{i}}$ seems to
converge well towards the system center as shown in
Fig.~\ref{fig:Szi}(c,e). Let us then look over what happens
if the edge spins at $i=1$, $i=L^2$ are pinned. The Friedel
oscillation for this case is shown in Fig.~\ref{fig:Szi}(b).
We find that the amplitude near the pinned spins is visibly
smaller than that around the system center. Accordingly, the
amplitude tends to be rather small for fixed $L$ as in
Fig.~\ref{fig:Szi}(d,f). Hence perhaps, the order parameter
might be underestimated or the convergence to the thermodynamic
limit could be slow. Since the outer $L-1$ sites for each
edge lose the original bond connections by cutting the
projected 1D periodic chain, it is more reasonable to regard
the outer $2(L-1)$ spins in total as `edged-spin group'.
Thus, the pinnings for the inner spins placed at $i=L$,
$i=L^2-L+1$ would be a good choice.

\begin{figure}[tb]
	\centering
	\includegraphics[width=1.0\columnwidth]{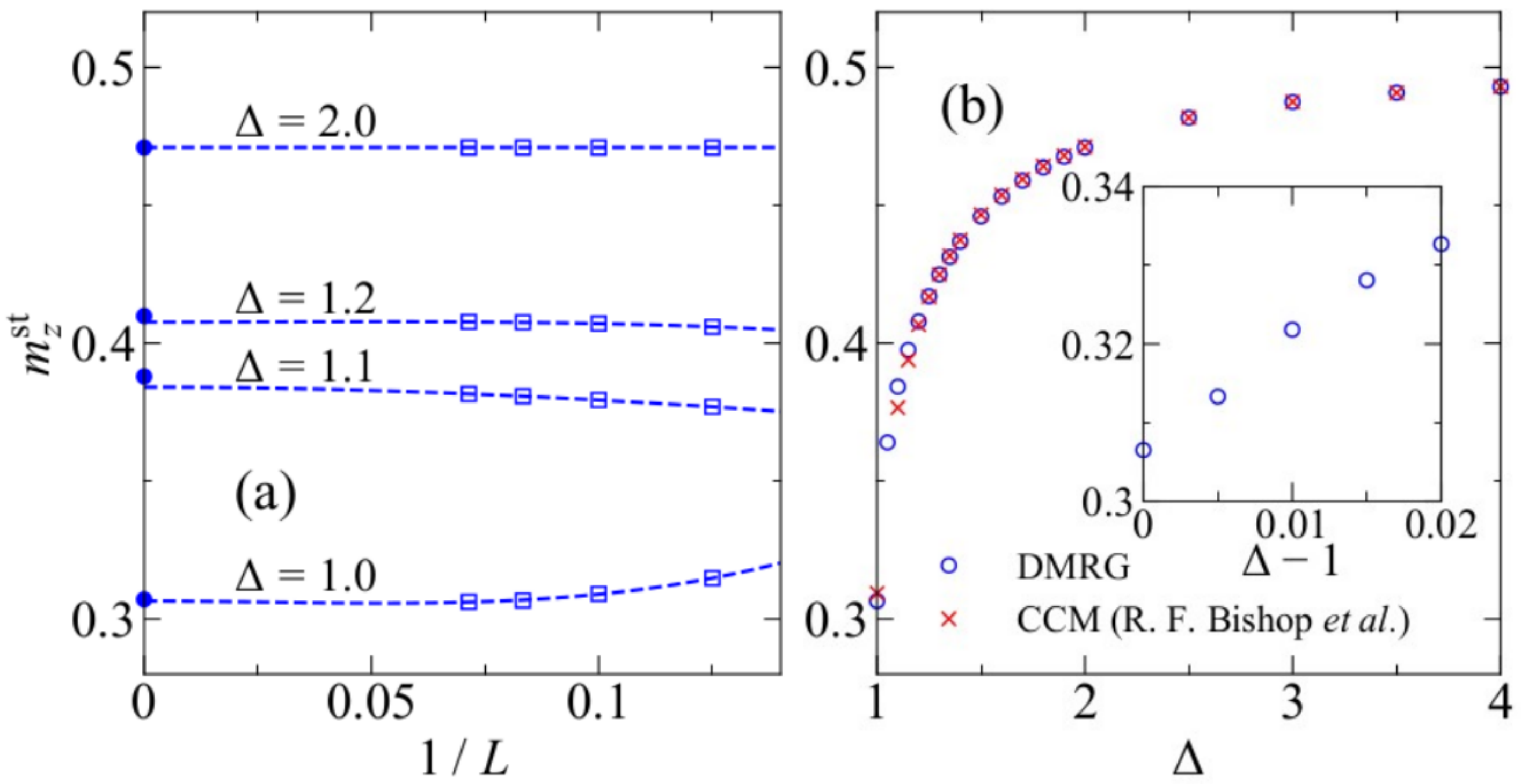}
    \caption{(Color online) (a) Finite-size scaling analysis of
	$m_{z}^{\rm{st}}$ for $S=1/2$ and $\Delta \ge 1$. Solid
	circles at $1/L=0$ are our previous results using periodic
	chains~\cite{Kadosawa2022}.
	(b) Extrapolated values of $m_{z}^{\mathrm{st}}$ as a
	function of $\Delta$. The results of CCM~\cite{BISHOP2017}
	are shown by red crosses. The inset shows
	$m_{z}^{\mathrm{st}}$ as a function of $\Delta - 1$ for a
	narrow range near $\Delta = 1$.}
	\label{fig:Delta_mzst}
\end{figure}

We begin by discussing the performance of our method at
$\Delta=1$ between the easy-axis N\'eel and XY phases for
$S = 1/2$. Since quantum fluctuations are largest and
entanglement range is maximized in this case, it is rather
difficult to numerically estimate the magnitude of
magnetization for the thermodynamic limit. Actually,
its accurate estimation had been a longstanding problem until
2010s~\cite{White2007,Sandvik2010}. Therefore, this is a good
benchmark to evaluate our method. The direction of magnetization
is now arbitrary. So, assuming it parallel to the
$z$-axis, we calculate $m_{z}^{\mathrm{st}}$. In this way,
we can restrict the spin configurations to a subspace with
$S_z^{\rm{tot}} = \sum_iS^z_i=0$.
In Fig.~\ref{fig:Delta_mzst}(a) we perform
finite-size scaling analysis of $m_{z}^{\mathrm{st}}$, where
open chains with length up to $N = L^2 = 196$ sites are studied.
The convergence of $m_{z}^{\mathrm{st}}$ with $1/L$ seems to
be fast enough to perform a reliable scaling. By fitting with
$m_{z}^{\mathrm{st}}(L)=m_{z}^{\mathrm{st}}+a/L^2+b/L^3$,
we obtain $m_{z}^{\mathrm{st}} = 0.3065$, which is in good
agreement with those by previous numerical estimations:
DMRG ($m_{z}^{\mathrm{st}}=0.3067$)~\cite{White2007},
QMC ($m_{z}^{\mathrm{st}}=0.30743$)~\cite{Sandvik2010},
and CCM ($m_{z}^{\mathrm{st}}=0.3093$)~\cite{BISHOP2017}.
We also note that this value is only slightly smaller than our
previous estimation using periodic chains
($m_{z}^{\mathrm{st}} = 0.3071 \pm 0.0005$)~\cite{Kadosawa2022}.

\begin{table}[bt]
	\caption{Extrapolated values of $m_{z}^{\mathrm{st}}$ in
	the easy-axis N\'eel phase ($\Delta\ge1$) for $S=1/2$.
	Results obtained by DMRG and CCM methods are compared:
	$\delta=m_{z}^{\mathrm{st}}({\rm CCM})-m_{z}^{\mathrm{st}}({\rm DMRG})$.
	}
	\begin{center}
		\begin{tabular}{wc{1.4cm}wc{1.8cm}wc{1.8cm}wc{1.8cm}}
			\hline
			\hline
			$\Delta$ &  $m_{z}^{\mathrm{st}}$(DMRG)  & $m_{z}^{\mathrm{st}}$(CCM) &  $\delta$ \\
			\hline
			1.0  &   0.3065  &  0.3093  &  0.0028 \\
			1.2  &   0.4079  &  0.4067  & $-$0.0012 \\
			1.5  &   0.4459  &  0.4466  &  0.0007 \\
			2.0  &   0.4709  &  0.4712  &  0.0003 \\
			3.0  &   0.4874  &  0.4875  &  0.0001\\
			4.0  &   0.4930  &  0.4930  & $<$0.0001\\
			\hline
			\hline
		\end{tabular}
	\end{center}
	\label{Table:Deltage1}
\end{table}

The system is gapless at $\Delta=1$. While on the other hand,
the gap opens between the $S^z_{\rm tot}=0$ ground state
and $S^z_{\rm tot}=1$ excited state for $\Delta>1$. This
implies that the direction of staggered magnetization is
uniquely fixed parallel to the $z$-axis. Let us then estimate
$m_{z}^{\mathrm{st}}$ for $\Delta>1$. Finite-size scaling
analyses of $m_{z}^{\mathrm{st}}$ for some $\Delta$ values
are shown in Fig.~\ref{fig:Delta_mzst}(a). For larger $\Delta$
the size-dependence of $m_{z}^{\mathrm{st}}$ is smaller as
expected from the fact that entanglement range is reduced by
the reduction of quantum fluctuations.
The extrapolated values of $m_{z}^{\mathrm{st}}$ are plotted
as a function of $\Delta$ in Fig.~\ref{fig:Delta_mzst}(b).
We can see that $m_{z}^{\mathrm{st}}$ converges rapidly to
the Ising value $1/2$ with $\Delta$. For comparison, recent
data from CCM method~\cite{BISHOP2017} are also shown.
The agreement seems to be overall good. The values of
$m_{z}^{\mathrm{st}}$ are particularly compared in
Table~\ref{Table:Deltage1}. The deviation becomes larger with
approaching $\Delta=1$. Nevertheless, even though the
uncertainties of the scaling to $L\to\infty$ in DMRG as well
as of the scaling to $m\to\infty$ in the so-called LSUB$_m$
level of approximation in CCM are maximal at $\Delta=1$, the
error of $m_{z}^{\mathrm{st}}$ is only $\sim0.9\%$. We also
mention the critical behavior of $m_{z}^{\mathrm{st}}$ near
$\Delta=1$. A singularity expressed by
$m_z^{\rm st}=\sum_{n=0}^{\infty}m_n(\Delta-1)^{n/2}$ was
predicted by SWT~\cite{Weihong1991-1,Weihong1991-2,Huse1988}.
However, as shown in the inset of Fig.~\ref{fig:Delta_mzst}(b),
we find that $m_{z}^{\mathrm{st}}$ is nearly proportional to
$\Delta-1$ in the range of $1\le\Delta\lesssim1.01$. This is
consistent with the result from CCM~\cite{BISHOP2017}.
To further demonstrate the accuracy of our method, we provide
a precise comparison of our large-$\Delta$ data with the result
obtained by SE for $1/\Delta$. By fitting our data for
$0 \le 1/\Delta \le 0.05$ with
$2m_{z}^{\mathrm{st}} = 1 + m_2/\Delta^2 + m_4/\Delta^4 + m_6/\Delta^6$, we obtain $m_2= -0.222222225$, $m_4= -0.0355542736$, and $m_6=-0.0189663810$.
These coefficients agree almost perfectly to the SE results:
$m_2=-2/9=-0.222222\dots$, $m_4=-8/255=-0.0355555\dots$,
and $m_6=-0.01894258$~\cite{Davis1960,Huse1988,Parrinello1974,Singh1989,Weihong1991-1}.

\begin{figure}[tb]
	\centering
	\includegraphics[width=1.0\columnwidth]{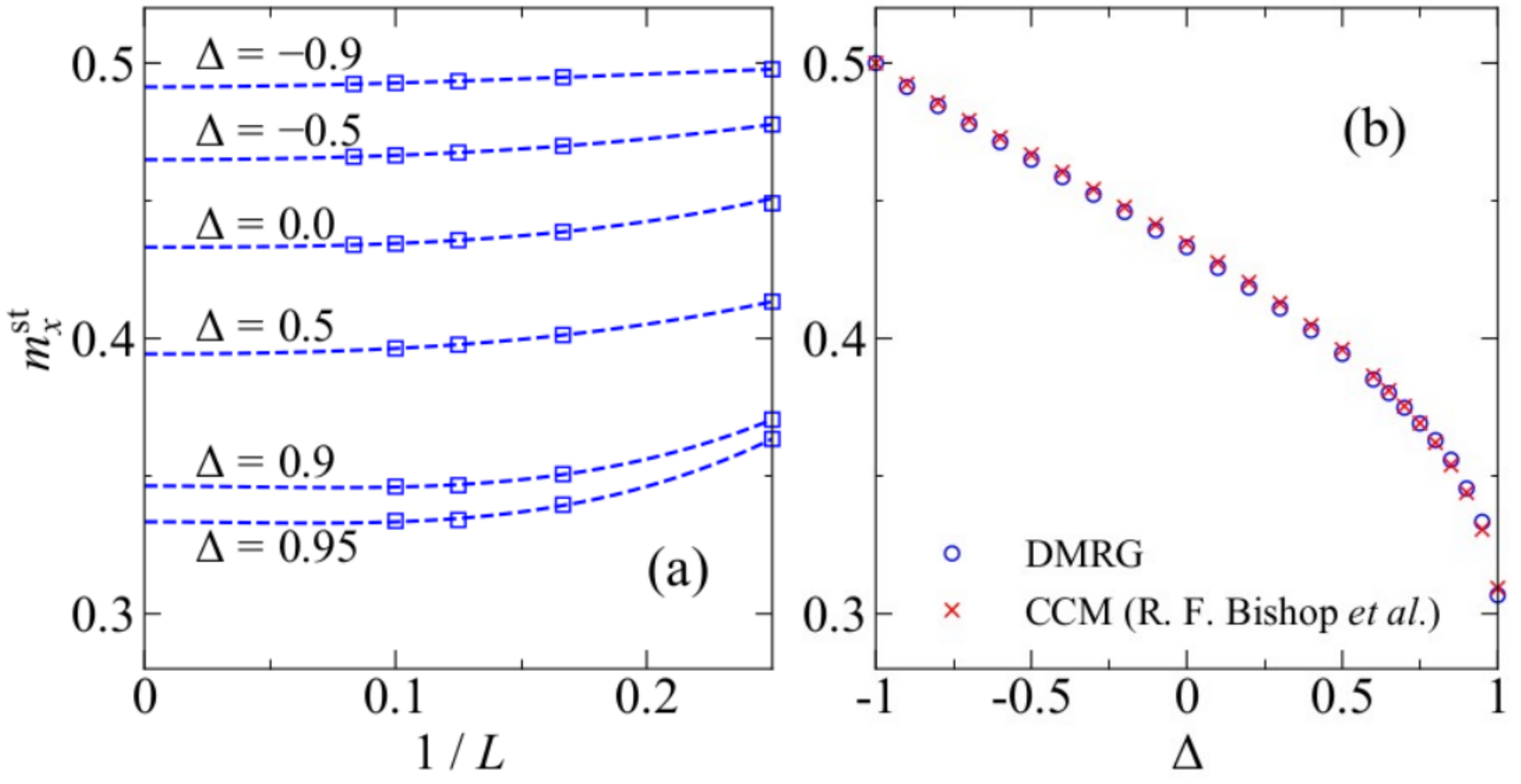}
    \caption{(Color online) (a) Finite-size scaling analysis of
	$m_{x}^{\mathrm{st}}$ for $S=1/2$ and $|\Delta| < 1$.
	(b) Extrapolated values of $m_{x}^{\mathrm{st}}$ as a
	function of $\Delta$. The results of CCM~\cite{BISHOP2017}
	are shown by red crosses.
		}
	\label{fig:Delta_mxst}
\end{figure}

\begin{table}[bt]
	\caption{Extrapolated values of $m_{x}^{\mathrm{st}}$ in
	the XY phase ($-1<\Delta\le1$) for $S=1/2$. Results
	obtained by DMRG and CCM methods are compared:
	$\delta=m_{x}^{\mathrm{st}}({\rm CCM})-m_{x}^{\mathrm{st}}({\rm DMRG})$.
	}
	\begin{center}
		\begin{tabular}{wc{1.4cm}wc{1.8cm}wc{1.8cm}wc{1.8cm}}
			\hline
			\hline
			$\Delta$ &  $m_{x}^{\mathrm{st}}$(DMRG)  & $m_{x}^{\mathrm{st}}$(CCM) &  $\delta$ \\
			\hline
			-1.0  &   0.5000  &  0.5000  &  0.0000 \\
			-0.8  &   0.4844  &  0.4856  &  0.0012 \\
			-0.5  &   0.4649  &  0.4667  &  0.0018 \\
			0.0  &   0.4331  &  0.4346  &  0.0015 \\
			0.5  &   0.3943  &  0.3960  &  0.0017 \\
			1.0  &   0.3065  &  0.3093  &  0.0028 \\
			\hline
			\hline
		\end{tabular}
	\end{center}
	\label{Table:DeltaXY}
\end{table}

Similarly, the magnitude of staggered magnetization in the
XY phase ($-1<\Delta< 1$) can be obtained. Since the
magnetization is parallel to some arbitrary direction in
the $xy$-plane, we here measure $m_{x}^{\mathrm{st}}$ with
two pinned spins $\braket{S^{x}_{L}}=1/2$ and $\braket{S^{x}_{L^2-L+1}}=-1/2$. In this case, the DMRG
calculations are a little more difficult than the above
estimations of $m_{z}^{\mathrm{st}}$ because $S_z^{\rm{tot}}$
is no longer conserved. Still, as shown below, sufficient
data points to perform a reliable finite-size scaling analysis
are available. Fig.~\ref{fig:Delta_mxst}(a) shows the
finite-size scaling analysis of $m_{x}^{\mathrm{st}}$ for
various $\Delta$ values in the XY phase, where open chains
with length up to $N = L^2 = 144$ sites are used. As expected,
the size-dependence of $m_{x}^{\mathrm{st}}$ becomes smaller
with approaching the classical limit ($\Delta=-1$) from
the isotropic point ($\Delta=1$).
In Fig.~\ref{fig:Delta_mxst}(b) the extrapolated values of
$m_{x}^{\mathrm{st}}$ are plotted as a function of $\Delta$.
For comparison, the results from CCM~\cite{BISHOP2017}
are also shown. We see that $m_{x}^{\mathrm{st}}$ increases
smoothly with decreasing $\Delta$ and approaches $1/2$ for
$\Delta \to -1$. This confirms that the quantum fluctuations
of this system are strongest at $\Delta=1$. The agreement with
CCM results looks overall good. In addition, the values of
$m_{x}^{\mathrm{st}}$ are compared for some $\Delta$ values in
Table~\ref{Table:DeltaXY}. The DMRG values are only slightly
smaller than the CCM ones for each $\Delta$.
This may be because that DMRG can incorporate
more quantum fluctuations compared to CCM. Furthermore,
we can find that our estimation of 
$m_{x}^{\mathrm{st}}=0.4331$ at the XY point ($\Delta=0$)
reasonably agrees to the previous estimations: 
$m_{x}^{\mathrm{st}} = 0.43548$ by SE~\cite{Hamer1991} and $m_{x}^{\mathrm{st}} = 0.437$ by QMC~\cite{Sandvik1999}.

So far, We have verified the efficiency of our proposed method
for $S=1/2$. In practice, there is no difficulty
involved in finite-size scaling of
$m_{\alpha}^{\mathrm{st}}$ ($\alpha=x$, $z$), so that 
the extrapolated values to the thermodynamic limit are quite 
accurate in the whole $\Delta$ region as far as has been
compared to the previous studies. We then apply our method to
estimate $m_{z}^{\mathrm{st}}$ for the high-$S$ cases from
$S=1$ to $S=6$, which is a challenging problem for numerical
calculations. 

\begin{figure}[tb]
	\centering
	\includegraphics[width=1\columnwidth]{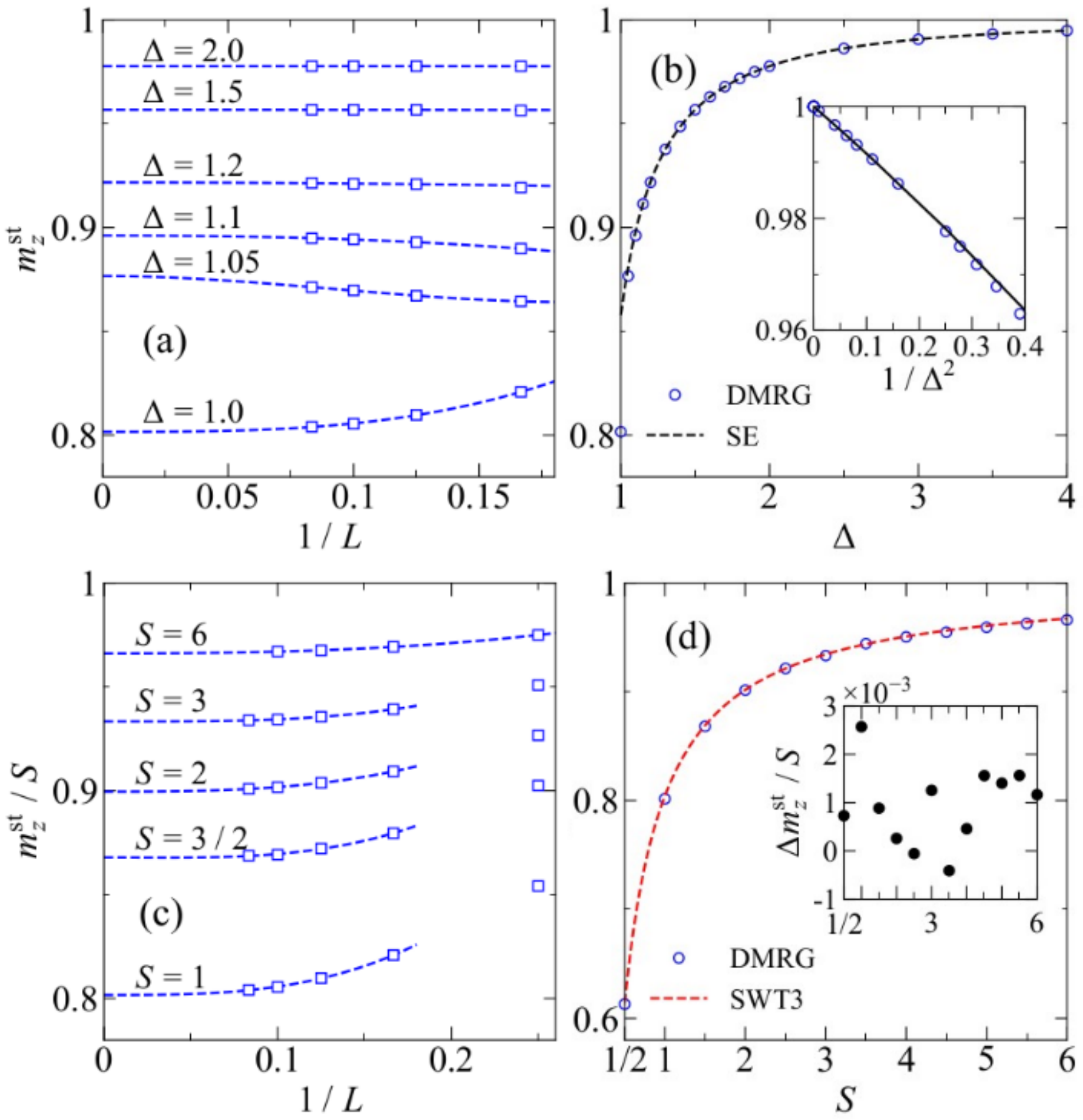}
    \caption{(Color online) (a) Finite-size scaling analysis of
	$m_{z}^{\mathrm{st}}$ and (b) extrapolated values of
	$m_{z}^{\mathrm{st}}$ as a function of $\Delta$ for $S=1$
	and $\Delta \ge 1$. Inset: $m_{z}^{\mathrm{st}}$ vs.
	$1/\Delta^2$ in the large $\Delta$ region. The dashed line
	shows the SE result~\cite{Singh1990,Weihong1991-1}.
	(c) Finite-size scaling analysis of $m_{z}^{\mathrm{st}}$
	and (d) extrapolated values of $m_{z}^{\mathrm{st}}$ as a
	function of $S$ at $\Delta=1$. The SWT
	results~\cite{Hamer1992,Igarashi1992,Canali1993} are shown
	by the red line. Inset: Difference between our DMRG and
	SWT results~\cite{Hamer1992,Igarashi1992,Canali1993}.}
	\label{fig:S_mzst}
\end{figure}

Let us start with the case of $S=1$. It is known that our system
for $S\ge1$ exhibits an N\'eel order for
$\Delta\ge-1$~\cite{Kubo1988,Ozeki1989}. As in the case of
$S=1/2$, we can calculate $m_{z}^{\mathrm{st}}$ for $\Delta\ge1$.
Finite-size scaling analyses of $m_{z}^{\mathrm{st}}$ are
performed in Fig.~\ref{fig:S_mzst}(a). The extrapolated values
of $m_{z}^{\mathrm{st}}$ are plotted as a function of $\Delta$
in Fig.~\ref{fig:S_mzst}(b). We obtain
$m_{z}^{\mathrm{st}}=0.8017$ at $\Delta=1$ in the thermodynamic
limit. This value is in good agreement with
$m_{z}^{\mathrm{st}}=0.802$
obtained by iPEPS~\cite{Niesen2017}. The SE
results~\cite{Singh1990,Weihong1991-1} are also shown for
comparison. By fitting our data for $0 \le 1/\Delta \le 0.03$
with $m_{z}^{\mathrm{st}} = 1 + m_2/\Delta^2 + m_4/\Delta^4 + m_6/\Delta^6$, we obtain $m_2=-0.081632653$, $m_4=-0.026959397$,
and $m_6=-0.013867377$. These values are in good agreement with
those from SE: $m_2=-4/49=-0.081632653\dots$, $m_4=-0.026959099$,
and $m_6=-0.0136997515$~\cite{Singh1990,Weihong1991-1}.

We next study the cases of $S>1$. Fig.~\ref{fig:S_mzst}(c) shows
the finite-size scaling analysis of $m_{z}^{\mathrm{st}}$ for
$S\ge1$, where open chains with length up to $N=L^2=144$ sites
are studied. As expected from the fact that the effect of
quantum fluctuations is weaker for larger $S$, the
size-dependence of $m_{z}^{\mathrm{st}}$ becomes smaller with
increasing $S$. However, the scaling analysis can be easily
done even for the smallest-$S$, i.e., $S=1$ case.
In Fig.~\ref{fig:S_mzst}(d) the extrapolated values of
$m_{z}^{\mathrm{st}}/S$ are plotted as a function of $S$.
We can clearly see a smooth convergence
as $m_{z}^{\mathrm{st}}/S\to1$ with approaching the classical
limit ($S=\infty$). The $S$-dependence of $m_{z}^{\mathrm{st}}/S$
has been estimated by SWT. The result up to third order is 
$m_{z}^{\mathrm{st}}/S = 1 - 0.1966019S^{-1} + 0.00087S^{-3} + O(S^{-4})$~\cite{Hamer1992,Igarashi1992,Canali1993}. This
expression may be expected to work well for a wide range of $S$
because the coefficients of higher order terms than $1/S$ are
very small. Actually, as shown in the inset of
Fig.~\ref{fig:S_mzst}(d), the difference for each $S$ value
between our results and the SWT estimations is always smaller
than $\Delta m_{z}^{\mathrm{st}}/S=3\times10^{-3}$. By fitting
our data points from $S=1/2$ to $S=6$ with
$m_{z}^{\mathrm{st}}/S = 1+m_{1}S^{-1}+m_{3}S^{-3}$, we obtain
$m_{1}=-0.19895398$ and $m_{3}=0.00136057$. These values are
reasonably close to those by SWT. Therefore, we confirm that
our method is applicable to the high-$S$ Heisenberg systems
at least up to $S=6$. We can also numerically confirm the
absence of Haldane-like state with spin-singlet pairs on every
bonds at $S=2$~\cite{Kubo1988,Ozeki1989}.

In conclusion, we proposed an efficient method to obtain a local
order parameter for 2D systems by DMRG using SBC. We demonstrated
the validity of our method by calculating staggered
magnetization of the $S=1/2$ XXZ square-lattice Heisenberg model
for the whole range of exchange anisotropy. As further
application, we extended our method to the higher spin cases
($1 \le S \le 6$). Although we investigated an order parameter
with $\bm{k}=(\pi,\pi)$ in this paper, ordered state with the
other $\bm{k}$ vectors can be considered by modifying the
application of SBC~\cite{Nakamura2021}.
Other examples of such SBC usage are given in
the Supplemental Material~\cite{SM}.
We also note that a similar procedure using the projected 1D chain
with periodic boundary conditions was suggested in our previous
study~\cite{Kadosawa2022}. However, the present method is much
more practical because larger clusters can be studied than the
previous method.
In order to further clarify the advantages
of SBC in DMRG simulation, finite-size scaling analysis for various
boundary conditions is discussed in the Supplemental Material~\cite{SM}.

\begin{acknowledgment}
We thank Ulrike Nitzsche for technical support.
This work was supported by the SFB 1143 of the Deutsche Forschungsgemeinschaft and by 
Grants-in-Aid for Scientific Research from JSPS (Projects No. JP20H01849, No. JP20K03769, and No. JP21J20604).
M.~K. acknowledges support from the JSPS Research Fellowship for Young Scientists.
\end{acknowledgment}


\begin{thebibliography}{99}
\bibitem{Heisenberg1928} W.~Heisenberg, Z. Phys. {\bf 49}, 619 (1928).
\bibitem{Lines1963} M.E.~Lines, Phys. Rev. {\bf 131}, 546 (1963).
\bibitem{Achiwa1969} N.~Achiwa, J. Phys. Soc. Jpn. {\bf 27}, 561 (1969).
\bibitem{Kosterlitz1973} J.M.~Kosterlitz and D.J.~Thouless, J. Phys. C: Solid State Phys. {\bf 6}, 1181 (1973).
\bibitem{Opherden2022} D.~Opherden, M.S.J.~Tepaske, F.~B\"artl, M.~Weber, M.M.~Turnbull, T.~Lancaster, S.J.~Blundell, M.~Baenitz, J.~Wosnitza, C.P.~Landee, R.~Moessner, D.J.~Luitz, H.~K\"uhne, arXiv:2209.11085.
\bibitem{Wu2021} Y.~Wu, J.~Xi, T.~Xiao, J.~Ferrando-Soria, Z.~Ouyang, Z.~Wang, S.~Luo, X.~Liu, and E.~Pardo, Inorg. Chem. Front. {\bf 8}, 5158 (2021).
\bibitem{Sonin2010} E.B.~Sonin, Advances in Physics {\bf 59}, 181 (2010).
\bibitem{Jang2021} T.-H.~Jang, S.-H.~Do, M.~Lee, H.~Wu, C.M.~Brown, A.D.~Christianson, S.-W.~Cheong, and J.-H.~Park, Phys. Rev. B {\bf 104}, 214434 (2021).
\bibitem{Do2022} S.-H.~Do, H.~Zhang, D.A.~Dahlbom, T.J.~Williams, V.O.~Garlea, T.~Hong, T.-H.~Jang, S.-W.~Cheong, J.-H.~Park, K.~Barros, C.D.~Batista, and A.D.~Christianson, arXiv:2205.11770
\bibitem{Lee2022} M.~Lee, R.~Schoenemann, H.~Zhang, D.~Dahlbom, T.-H.~Jang, S.-H.~Do, A.D.~Christianson, S.-W.~Cheong, J.-H.~Park, E.~Brosha, M.~Jaime, K.~Barros, C.D.~Batista, and V.S.~Zapf, arXiv:2210.14323
\bibitem{Vasilchikova2020} T.~Vasilchikova, V.~Nalbandyan, I.~Shukaev, H.-J.~Koo, M.-H.~Whangbo, A.~Lozitskiy, A.~Bogaychuk, V.~Kuzmin, M.~Tagirov, E.~Vavilova, A.~Vasiliev, and E.~Zvereva, Phys. Rev. B {\bf 101}, 054435 (2020).
\bibitem{Kuchugura2019} M.D.~Kuchugura, A.I.~Kurbakov, E.A.~Zvereva, T.M.~Vasilchikova, G.V.~Raganyan, A.N.~Vasiliev, V.A.~Barchukf, and V.B.~Nalbandyan, Dalton Trans. {\bf 48}, 17070 (2019).
\bibitem{Baenitz2021} M.~Baenitz, M.M.~Piva, S.Luther, J.~Sichelschmidt, K.M.~Ranjith, H.~Dawczak-D\c{e}bicki, M.O.~Ajeesh, S.-J.~Kim, G.~Siemann, C.~Bigi, P.~Manuel, D.~Khalyavin, D.A.~Sokolov, P.~Mokhtari, H.~Zhang, H.~Yasuoka, P.D.C.~King,G.~Vinai, V.~Polewczyk, P.~Torelli, J.~Wosnitza, U.~Burkhardt, B.~Schmidt, H.~Rosner, S.~Wirth, H.~K\"uhne, M.~Nicklas, and M.~Schmidt, Phys. Rev. B {\bf 104}, 134410 (2021).
\bibitem{Liu2021} J.~Liu, B.~Liu, L.~Yuan, B.~Li, L.~Xie, X.~Chen, H.~Zhang, D.~Xu, W.~Tong, J.~Wang, and Y.~Li, New J. Phys. {\bf 23}, 033040 (2021).
\bibitem{Rawl2019} R.~Rawl, L.~Ge, Z.~Lu, Z.~Evenson, C.R.D.~Cruz, Q.~Huang, M.~Lee, E.S.~Choi, M.~Mourigal, H.D.~Zhou, and J.~Ma, Phys. Rev. Materials {\bf 3}, 054412 (2019).
\bibitem{Yoshida2020} H.K.~Yoshida, M.~Matsuda, M.B.~Stone, C.R.~dela Cruz, T.~Furubayashi, M.~Onoda, E.~Takayama-Muromachi, and M.~Isobe, Phys. Rev. Research {\bf 2}, 043211 (2020).
\bibitem{Kim2020} C.~Kim, J.~Jeong, P.~Park, T.~Masuda, S.~Asai, S.~Itoh, H.-S.~Kim, A.~Wildes, and J.-G.~Park, Phys. Rev. B {\bf 102}, 184429 (2020).
\bibitem{Goto2018} M.~Goto, H.~Ueda, C.~Michioka, A.~Matsuo, K.~Kindo, K.~Sugawara, S.~Kobayashi, N.~Katayama, H.~Sawa, and K.~Yoshimura, Phys. Rev. B {\bf 97}, 224421 (2018).
\bibitem{Kermarrec2021} E.~Kermarrec, R.~Kumar, G.~Bernard, R.~Hénaff, P.~Mendels, F.~Bert, P.L.~Paulose, B.K.~Hazra, and B.~Koteswararao, Phys. Rev. Lett. {\bf 127}, 157202 (2021).
\bibitem{Ferrenti2022} A.M.~Ferrenti, V.~Meschke, S.~Ghosh, J.~Davis, N.~Drichko, E.S.~Toberer, T.M.~McQueen, J. Solid State Chem. {\bf 317}, 123620 (2022).
\bibitem{Sano2018} R.~Sano, Y.~Kato, and Y.~Motome, Phys. Rev. B {\bf 97}, 014408 (2018).
\bibitem{Lee2020} I.~Lee, F.G.~Utermohlen, D.~Weber, K.~Hwang, C.~Zhang, J.~van Tol, J.E.~Goldberger, N.~Trivedi, and P.C.~Hammel, Phys. Rev. Lett. {\bf 124}, 017201 (2020).
\bibitem{Xu2020} C.~Xu, J.~Feng, M.~Kawamura, Y.~Yamaji, Y.~Nahas, S.~Prokhorenko, Y.~Qi, H.~Xiang, and L.~Bellaiche, Phys. Rev. Lett. {\bf 124}, 087205 (2020).
\bibitem{White2007} S.R.~White and A.L.~Chernyshev, Phys. Rev. Lett. {\bf 99}, 127004 (2007).
\bibitem{Sandvik1999} A.W.~Sandvik and C.J.~Hamer, Phys. Rev. B {\bf 60}, 6588 (1999).
\bibitem{Sandvik2010} A.W.~Sandvik and H.G.~Evertz, Phys. Rev. B {\bf 82}, 024407 (2010).
\bibitem{Hamer1992} C.J.~Hamer, Z.~Weihong, and P.~Arndt, Phys. Rev. B {\bf 46}, 6276 (1992).
\bibitem{Igarashi1992} J.-i.~Igarashi, Phys. Rev. B {\bf 46}, 10763 (1992).
\bibitem{Canali1993} C.M.~Canali and M.~Wallin, Phys. Rev. B {\bf 48}, 3264 (1993).
\bibitem{Davis1960} H.L.~Davis, Phys. Rev. {\bf 120}, 789 (1960).
\bibitem{Huse1988} D.A.~Huse, Phys. Rev. B {\bf 37}, 2380 (1988).
\bibitem{Parrinello1974} M.~Parrinello and T.~Arai, Phys. Rev. B {\bf 10}, 265 (1974).
\bibitem{Singh1989} R.R.P.~Singh, Phys. Rev. B {\bf 39}, 9760 (1989).
\bibitem{Singh1990} R.R.P.~Singh, Phys. Rev. B {\bf 41}, 4873 (1990).
\bibitem{Weihong1991-1} Z.~Weihong, J.~Oitmaa, and C.J.~Hamer, Phys. Rev. B {\bf 43}, 8321 (1991).
\bibitem{Weihong1991-2} Z.~Weihong, J.~Oitmaa, and C.J.~Hamer, Phys. Rev. B {\bf 44}, 11869 (1991).
\bibitem{Hamer1991} C.J.~Hamer, J.~Oitmaa, and Z.~Weihong, Phys. Rev. B {\bf 43}, 10789 (1991).
\bibitem{BISHOP2017} R.F.~Bishop, P.H.Y.~Li, R.~Zinke, R.~Darradi, J.~Richter, D.J.J.~Farnell, and J.~Schulenburg, J. Magn. Magn. Mater. {\bf 428}, 178 (2017).
\bibitem{Kubo1988} K.~Kubo and T.~Kishi, Phys. Rev. Lett. {\bf 61}, 2585 (1988).
\bibitem{Viswanath1994} V.S.~Viswanath, S.~Zhang, J.~Stolze, and G.~M\"uller, Phys. Rev. B {\bf 49}, 9702 (1994).
\bibitem{Yunoki2002} S.~Yunoki, Phys. Rev. B {\bf 65}, 092402 (2002).
\bibitem{Braiorr-Orrs2019} B.~Braiorr-Orrs, M.~Weyrauch, and M.V.~Rakov, Ukrainian Journal of Physics {\bf 61}, 613 (2019).
\bibitem{SM} (Supplemental material) Exact wavefunctions of the ground state at $\Delta=-1$, detailed data on the accuracy of our DMRG calculations as well as the advantages of SBC in DMRG against various boundary conditions are provieded online.
\bibitem{Nakamura2021} M.~Nakamura, S.~Masuda, and S.~Nishimoto, Phys. Rev. B {\bf 104}, L121114 (2021).
\bibitem{Kadosawa2022} M.~Kadosawa, M.~Nakamura, Y.~Ohta, and S.~Nishimoto, arXiv:2205.15775 (2022).
\bibitem{White1992} S.R.~White, Phys. Rev. Lett. {\bf 69}, 2863 (1992).
\bibitem{Ozeki1989} Y.~Ozeki, H.~Nishimori, and Y.~Tomita, J. Phys. Soc. Jpn. {\bf 58}, 82 (1989).
\bibitem{Niesen2017} I.~Niesen and P.~Corboz, Phys. Rev. B {\bf 95}, 180404(R) (2017).

\end{thebibliography}
\end{document}


\title{Supplemental Material for\\ ``Study of Staggered Magnetization in the Spin-$S$ Square-Lattice Heisenberg Model Using Spiral Boundary Conditions''}

\author{Masahiro Kadosawa$^1$, Masaaki Nakamura$^2$, Yukinori Ohta$^1$, and Satoshi Nishimoto$^{3,4}$}

\address{
$^1$ Department of Physics, Chiba University, Chiba 263-8522, Japan\\
$^2$ Department of Physics, Ehime University, Ehime 790-8577, Japan\\
$^3$ Department of Physics, Technical University Dresden, 01069 Dresden, Germany\\
$^4$ Institute for Theoretical Solid State Physics, IFW Dresden, 01069 Dresden, Germany
}

\date{\today}

\maketitle

\section{I. Exact ground state at $\Delta=-1$}

As mentioned in the main text, the system (1) exhibits
a first-order transition between easy-plane N\'eel (XY) and
ferromagnetic (FM) phases at $\Delta=-1$. At this critical
point, the ground state is two-fold degenerate with energy
$E_0=-2NS^2$. One of them for $\Delta<-1$ side is a simple
FM state expressed as 
\begin{equation}
	|\Psi_0(\mathrm{FM})\rangle=\frac{1}{\sqrt{2}}(|\Uparrow \rangle+|\Downarrow \rangle),
	\label{fm}
\end{equation}
where $|\Uparrow \rangle$ and $|\Downarrow \rangle$ denote
fully-polarized states towards the $z$ and $-z$ directions,
respectively. On the other hand, for $\Delta>-1$ side the spins
are antiferromagnetically aligned on the $xy$-plane. The
direction of magnetization is arbitrary. This state is
expressed as
\begin{equation}
	|\Psi_0(\mathrm{XY})\rangle=\sum_m \lambda_m |\psi_m \rangle,
	\label{xy}
\end{equation}
where $|\psi_m \rangle$ are bases restricted to 
$S^z_{\rm tot}=\sum_{i=1}^{L^2} \langle S^z_i \rangle=0$
subspace, and $m$ is summed over all possible combinations of
spin configurations. Two coefficients have a relation
$\lambda_m=(-1)^n\lambda_m'$ if a basis $|\psi_m \rangle$ is
obtained by $n$ exchange processes from another basis
$|\psi_m' \rangle$. In the following subsections, we show this
state specifically for the cases of $S=1/2$ and $S=1$.

\subsection{A. $S=1/2$}

In this case, a state for each site is represented by either
spin-up ($|\uparrow\rangle$) or spin-down ($|\downarrow\rangle$)
states. Because of $S^z_{\rm tot}=0$, an $N$-site lattice
system contains $\frac{N}{2}$ up and $\frac{N}{2}$ down
spins. Since there are only two states are allowed in each site,
the coefficient $|\lambda_m|$ takes a value of either
$\frac{1}{\sqrt{\cal P}}$ or $-\frac{1}{\sqrt{\cal P}}$.
The total number of spin configurations is given by
${\cal P}=\frac{N!}{(N/2)!(N/2)!}$.

\subsection{B. $S=1$}

In this case, a state for each site is any one of spin-up
($|\uparrow\rangle$), spin-down ($|\downarrow\rangle$), and
spin-0 ($|0\rangle$) states. Because of $S^z_{\rm tot}=0$,
the numbers of spin-up and spin-down sites should be the same.
By setting them to be $r/2$, the number of spin-0 sites is
$N-r$ for an $N$-site lattice system. The absolute value of
the coefficient depends on $r$ like $|\lambda_m|=2^{N-2r}|a|$.
The value of $a$ is determined by a condition
\begin{equation}
|a|^2\sum_{r=0}^{N/2} {}_N C_{2r} {}_{2r} C_r 2^{N-2r}=1.
	\label{coefS1}
\end{equation}

\newpage

\section{II. Comparison with series expansion of $1/\Delta$ for $S=1/2$}

\begin{figure}[h]
	\includegraphics[width=0.5\linewidth]{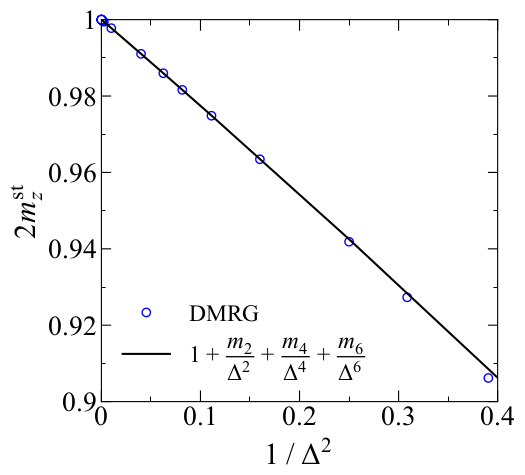}
	\caption{DMRG results for magnitude of the staggered
	magnetization of the $S=1/2$ square-lattice Heisenberg model
	as a function of $1/\Delta^2$. The solid line shows fitting
	by a polynomial function
	$2m_z^{\rm st}=1+m_2/\Delta^2+m_4/\Delta^4+m_6/\Delta^6$.
	}
	\label{fig:ek_SM}
\end{figure}

In Fig.~\ref{fig:ek_SM} the DMRG results for magnitude of the
staggered magnetization of the $S=1/2$ square-lattice Heisenberg
model are plotted as a function of $1/\Delta^2$. Our data in the
large-$\Delta$ region ($0\le1/\Delta\le0.05$) can be fitted by
$2m_z^{\rm st}=1+m_2/\Delta^2+m_4/\Delta^4+m_6/\Delta^6$ with
$m_2=-0.222222225$, $m_4=-0.0355542736$, and
$m_6=-0.0189663810$. This agrees well with the series expansions
$2m_z^{\rm st}=1-(2/9)/\Delta^2-(8/225)/\Delta^4-0.01894258/\Delta^6+{\cal O}(1/\Delta^8)$ [$2/9=0.22222222\dots$, $8/225=0.0355555\dots$]~\cite{Weihong1991-1}.

\newpage

\section{III. Comparison with series expansion of $1/S$}

\begin{figure}[h]
	\includegraphics[width=0.5\linewidth]{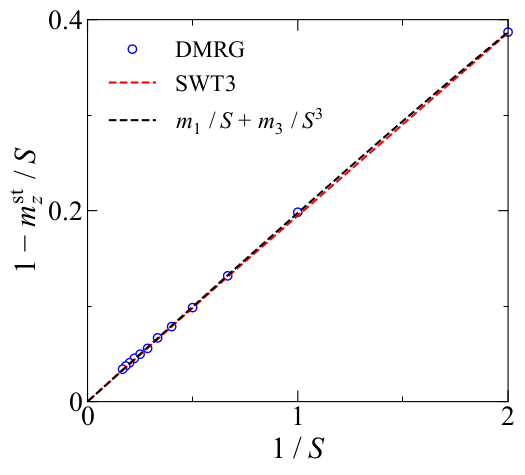}
	\caption{DMRG results for magnitude of the staggered
	magnetization of the square-lattice isotropic Heisenberg
	model as a function of $1/S$. The red and black dotted
	lines are SWT results and fitting by a polynomial function
	$m_{z}^{\mathrm{st}}/S = 1+m_{1}S^{-1}+m_{3}S^{-3}$,
	respectively.
	}
	\label{fig:anothersbc_SM}
\end{figure}

The $S$-dependence of $m_{z}^{\mathrm{st}}/S$ has been estimated
by spin wave theory (SWT). The result up to third order is
$m_{z}^{\mathrm{st}}/S = 1 - 0.1966019S^{-1} + 0.00087S^{-3} + O(S^{-4})$~\cite{Hamer1992,Igarashi1992,Canali1993}. This
expression may be expected to work well for a wide range of $S$
because the coefficients of higher order terms than $1/S$ are
very small. As shown in Fig.~\ref{fig:anothersbc_SM}, we can see
a good agreement between our results and SWT. By fitting our
data points from $S=1/2$ to $S=6$ with
$m_{z}^{\mathrm{st}}/S = 1+m_{1}S^{-1}+m_{3}S^{-3}$, we obtain
$m_{1}=-0.19895398$ and $m_{3}=0.00136057$. These values are
reasonably close to those by SWT.

\newpage

\section{IV. Flexibility of spiral boundary conditions}

\begin{figure}[h]
	\includegraphics[width=0.8\linewidth]{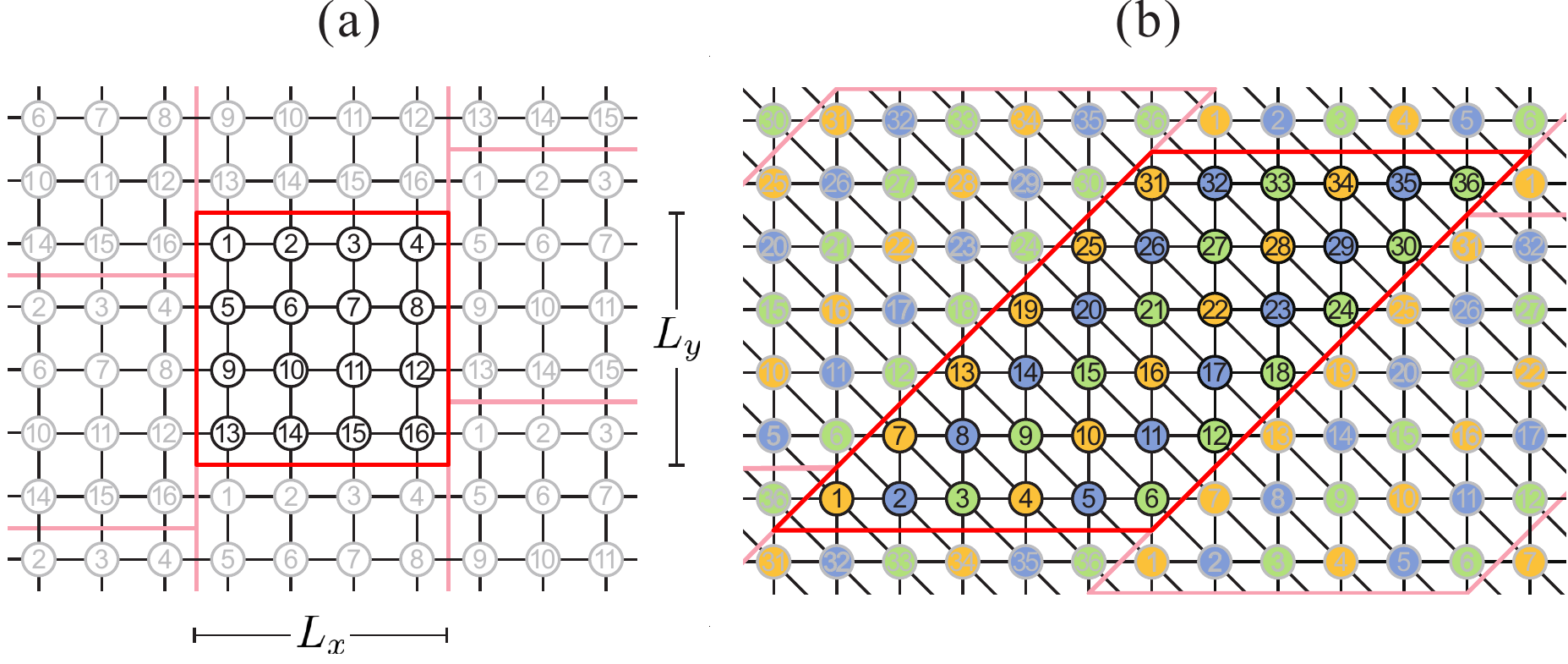}
	\caption{(a) 2D square-lattice cluster where a region framed by red line is the original cluster. This application of SBC is consistent with an ordered state with modulation of $\bm{k}=(\pi,0)$. (b) 2D triangular-lattice cluster where a region framed by red line is the original cluster. This application of SBC is consistent with an ordered state with modulation of $\bm{k}=(2\pi/3,2\pi/3)$. The sites belonging to different sublattices are color-coded.
	}
	\label{fig:anothersbc_SM}
\end{figure}

As noted in the main text, the ordering vector can be controlled by modifying
the application way of SBC~\cite{Nakamura2021}. The way used for 2D square lattice in the main
text is appropriate to study a state with modulation of $\bm{k}=(\pi,\pi)$,
like a N\'eel state. However, if we study a state with modulation of
$\bm{k}=(\pi,0)$, it is more convenient to use another application way of SBC.
It is shown in Fig.~\ref{fig:anothersbc_SM}(a). In this way, the original 2D
square-lattice cluster with $L_x \times L_y$ sites can be exactly mapped onto
a 1D chain with nearest- and $L_x$-th neighbor hopping integrals, where the translation symmetry is also preserved. The Hamiltonian is written as
\begin{equation}
	{\cal H}_{0,\rm sq} =-t \sum_\sigma\sum_{i=1}^{L_xL_y}(c^\dagger_{i,\sigma}c_{i+1,\sigma}+c^\dagger_{i,\sigma}c_{i+L_x,\sigma}+{\rm H.c.}),
\end{equation}
and its Fourier transform leads to
\begin{equation}
	{\cal H}_{{\rm sq},K} =-2t \sum_{K,\sigma}(\cos K + \cos L_xK)c^\dagger_{K,\sigma}c_{K,\sigma},
\end{equation}
where $n=n_x+L_xn_y=0,1,\dots,L_xL_y-1)$. It is interesting that both $\bm{k}=(\pi,\pi)$ in Fig.~1(a) of the main text and 
$\bm{k}=(\pi,0)$ in Fig.~\ref{fig:anothersbc_SM}(a) are 
represented by a $k=\pi$ modulation in the projected 1D chain.

As another example, let us consider a state with modulation of
$\bm{k}=(2\pi/3,2\pi/3)$. This corresponds to a 120$^\circ$
structure in the triangular-lattice Heisenberg model. Similarly,
this state can be represented by 1D chain with translation symmetry,
as shown in Fig.~\ref{fig:anothersbc_SM}(b). For a cluster with
$L_x \times L_y$ sites, the 2D system can be exactly mapped onto
a 1D chain with nearest-, $(L_x-2)$-th, and $(L_x-1)$-th neighbor bonds.

As shown above, the modulation of cluster can be flexibly controlled.
In the case of incommensurate order, it would be able to make
a convergence of physical quantity to the thermodynamic limit faster
by realizing a possibly closest modulation to the incommensurate vector.

\newpage

\section{V. Correlation functions for various boundary conditions}

\begin{figure}[h]
	\includegraphics[width=0.8\linewidth]{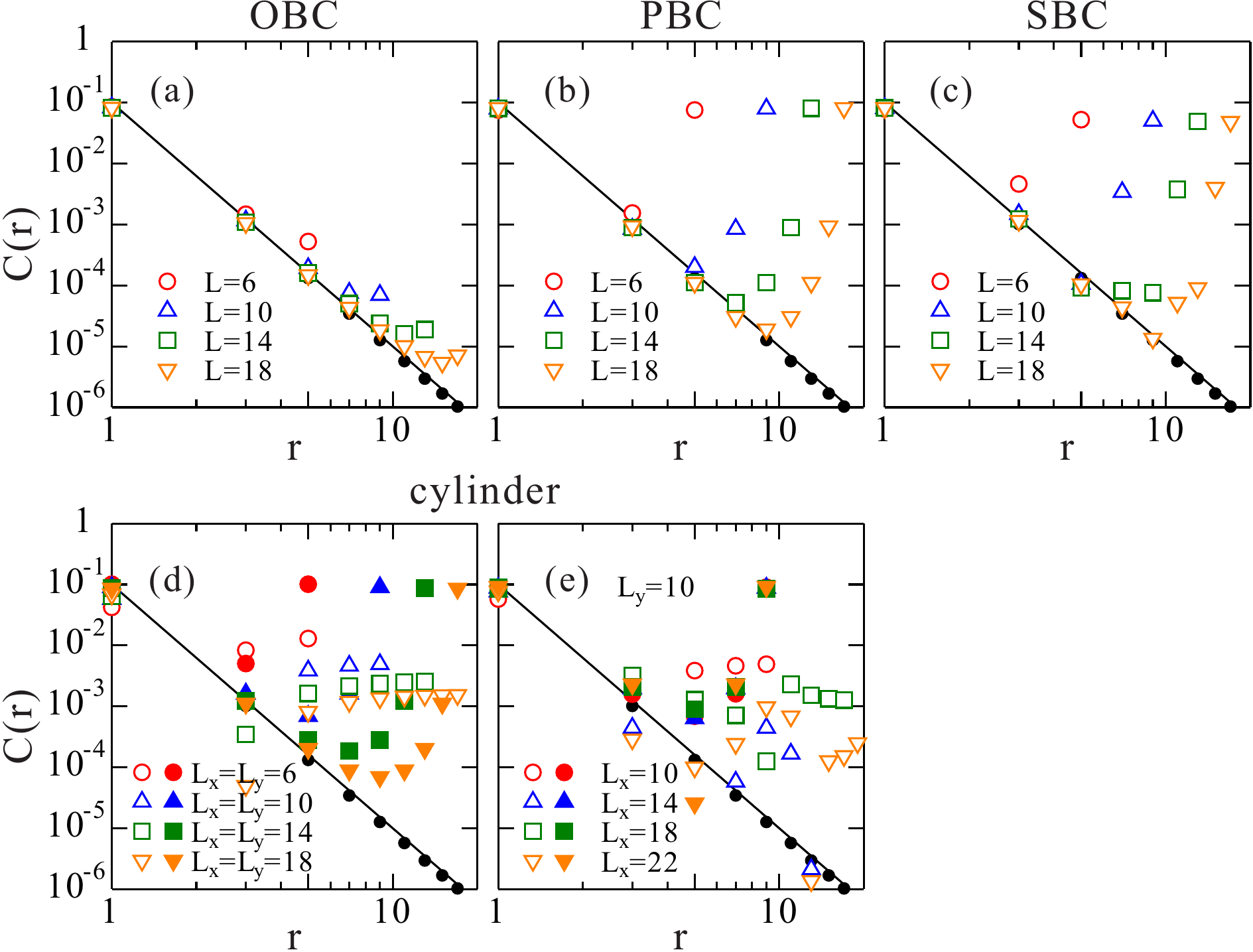}
	\caption{Density-density correlation functions $C(r)$ for non-interacting
		fermions on a 2D square lattice at half filling as a function of
		distance $r$ along bond direction, namely, $(1,0)$ or $(0,1)$, with
		(a) OBC, (b) PBC, (c) SBC, and (d,e) cylinder. For cylinder,
		the values for both $(1,0)$ and $(0,1)$ directions are shown.
		As a reference data, the result for very large system with $100\times100$ OBC cluster is plotted by black dots. As expected,
		it decays like $C(r) \propto 1/r^4$ (black line).
	}
	\label{fig:U0corr_SM}
\end{figure}

It is interesting to see how correlation functions are affected by boundary
conditions. To investigate their long-range behavior, we consider the
density-density correlation functions $C(r)=1-\langle n_i n_{i+r} \rangle$
of non-interacting fermions on a 2D square lattice because they can be
exactly obtained. In Fig.~\ref{fig:U0corr_SM} we plot
$C(r)=1-\langle n_i n_{i+r} \rangle$ as a function of distance $r$
along bond direction, namely, $(1,0)$ or $(0,1)$, at half filling for
various boundary conditions. We find that OBC, PBC, and SBC can provide
equally good results: For example, the data points up to $r\sim9$ are
almost on the analytical line $C(r) \propto 1/r^4$ for the systems with
$L=18$. On the other hand, the result with cylindrical cluster is obviously
worst, namely, the deviation from $C(r) \propto 1/r^4$ is largest.
As is well known~\cite{Sandvik}, the result seems to be improved
by increasing the system size with keeping the ratio between system length
and circumference. Nevertheless, the behavior $C(r) \propto 1/r^4$
can be confirmed only up to $r\sim5$ in the case of $L_x=L_y=18$
[see Fig.~\ref{fig:U0corr_SM}(d)]. It is because that cylindrical boundary
conditions impose a kind of spatial anisotropy on the original isotropic
2D square-lattice cluster. Furthermore, surprisingly, the result cannot
be improved by increasing the system length if the circumference is kept
[Fig.~\ref{fig:U0corr_SM}(e)].

As discussed above, we may choose either OBC or SBC when correlation functions
for 2D system are studied. However, a 2D system under OBC can be easily
disturbed by Friedel oscillations once it is doped or any interaction is
switched on. Therefore, SBC could be a first choice for DMRG study of correlation functions for 2D systems.

\newpage

\section{VI. Size scaling of ground-state energy for various boundary conditions}

\begin{figure}[h]
	\includegraphics[width=0.8\linewidth]{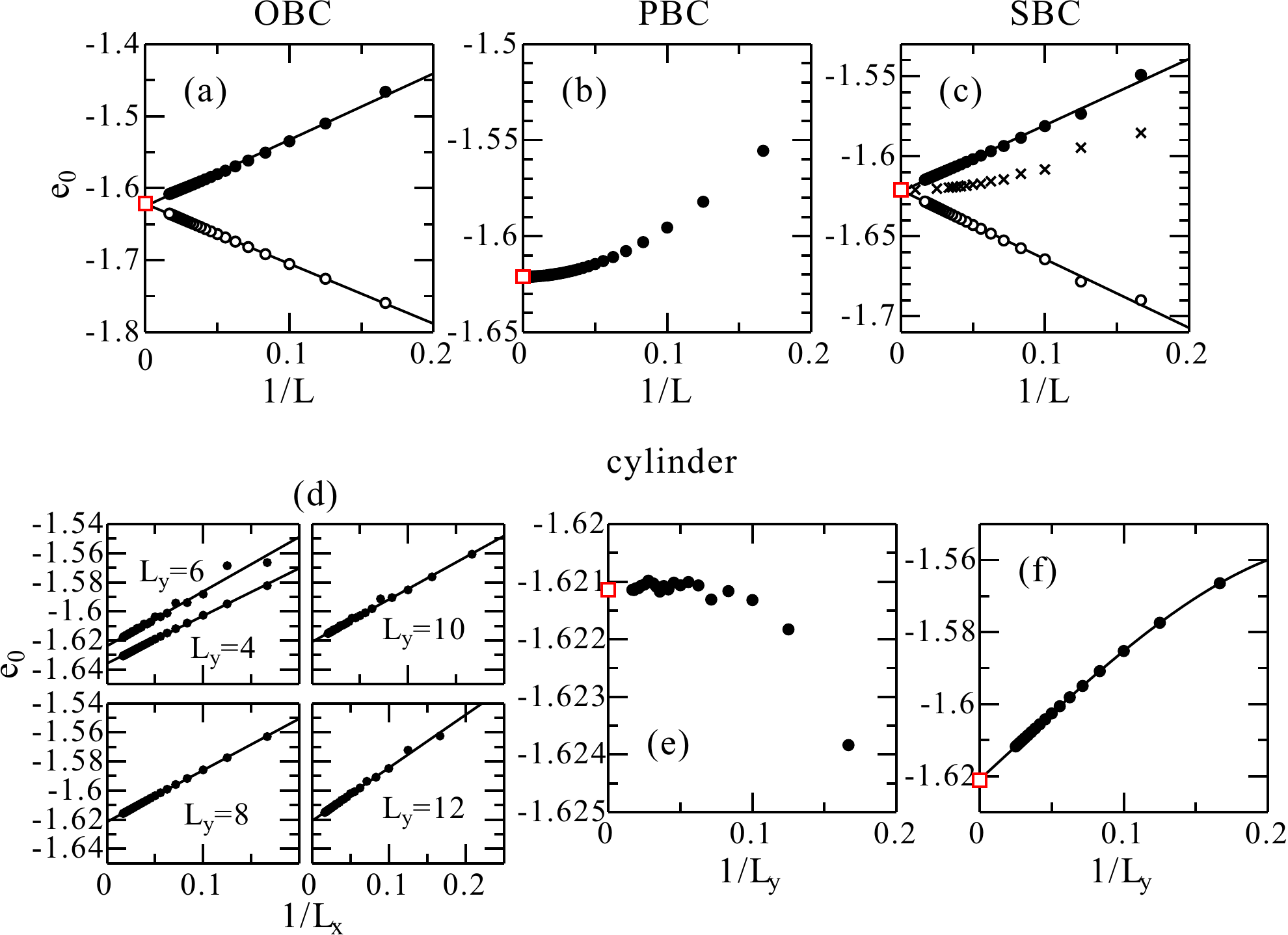}
	\caption{Finite-size-scaling analysis of the ground-state energy for
		non-interacting fermions on a 2D square lattice at half filling
		with (a) OBC, (b) PBC, (c) SBC, and (d-f) cylinder. The exact
		ground-state energy in the thermodynamic limit, $e_0=-16/\pi^2$,
		is denoted by red square. The solid and open circles denote the
		ground state energy per site and per two bonds, respectively.
		For cylinder, scaling analyses along system length $L_x$ for given
		circumferences $L_y$ are performed in (d), and then scaling analysis
		along circumference direction is performed in (e), where the
		extrapolated values obtained in (d) are used. Also, scaling analysis
		with keeping $L_x=L_y$ is performed in (f).
	}
	\label{fig:U0energy_SM}
\end{figure}

We here look into finite-size-scaling behavior of the ground-state energy
for each of boundary conditions. In Fig.~\ref{fig:U0energy_SM} we perform
finite-size-scaling analyses of the ground-state energy for non-interacting
fermions on a 2D square lattice at half filling with various boundary
conditions. The scaling behavior for OBC
is straightforward [Fig.~\ref{fig:U0energy_SM}(a)], which is similar to
that for the case of 1D projected open chain under SBC
[Fig.~\ref{fig:U0energy_SM}(c)]. On the other hand, the ground-state energy
approaches to its thermodynamic-limit value as $1/L^2$ for PBC. If we use
DMRG, the ground-state energies can be calculated only for $L\lesssim10$
with 2D periodic clusters. Thus, the finite-size-scaling analysis of the
ground-state energy would be difficult for PBC.

When we use cylinder, finite-size-scaling analysis must be performed along
two orientations, e.g., $x$- and $y$-directions. First, the scaling analyses
along system length $L_x$ for given circumferences $L_y$ are performed to
obtain the $L_x\to\infty$ values of the ground-state energy for each $L_y$
[Fig.~\ref{fig:U0energy_SM}(d)]. Next, with using the obtained $L_x\to\infty$
values further scaling analysis along circumference direction $L_y$ is
performed [Fig.~\ref{fig:U0energy_SM}(e)]. Similarly to the case of PBC,
the scaling behavior looks like $e_0=-16/\pi^2+\alpha/L_y^\beta$
($\alpha<0$, $\beta>2$). Therefore, accurate estimation of the ground-state
energy for 2D itinerant systems using DMRG with cylindrical clusters may
be not very practical, considering that there would be some uncertainty
in the extrapolation to $L_x\to\infty$ and the extrapolated values can be
obtained only for $L_y\lesssim10$. We also perform a scaling analysis with
keeping $L_x=L_y$ in Fig.~\ref{fig:U0energy_SM}(f). The scaling behavior
looks simple but deviates from a linear function like for OBC and open chain
under SBC.

For the above reason, OBC or SBC would be a best choice for estimating the
ground-state energy of 2D systems in the thermodynamic limit.

\newpage

\section{VII. Convergence check of DMRG calculations}

\begin{table}[tbh]
	\caption{Discarded weight $w_d$ and the ground-state energy $E_0(m)$
		of the half-filled square-lattice Hubbard model at $U/t=8$ as
		functions of system size $L \times L$ and the number of
		density-matrix eigenstates kept in the renormalization $m$,
		where SBC are applied. The ground-state energy at $m\to\infty$
		was obtained by a linear fit of $E_0(m)$ vs. $w_d$.
	}
	\begin{minipage}[t]{.48\textwidth}
		\begin{center}
			\begin{tabular}{wc{1.5cm}wc{1.8cm}wc{2cm}wc{3cm}}
				\multicolumn{4}{c}{$L=6$} \\
				\hline
				\hline
				$m$ &  $w_d$  & $E_0(m)$ &  $E_0(m)-E_0(\infty)$ \\
				\hline
				2000  &   3.432e-06  &  -17.79364790  &  0.01227902 \\
				4000  &   7.856e-07  &  -17.80326286  &  0.00266406 \\
				6000  &   3.052e-07  &  -17.80489349  &  0.00103344 \\
				8000  &   1.516e-07  &  -17.80541043  &  0.00051649 \\
				10000  &   8.614e-08  &  -17.80563244  &  0.00029448 \\
				12000  &   5.332e-08  &  -17.80574150  &  0.00018542 \\
				\hline
				$\infty$ &             &   -17.80592692  &            \\
				\hline
				\hline
			\end{tabular}
		\end{center}
	\end{minipage}
	\hfill
	\begin{minipage}[t]{.48\textwidth}
		\begin{center}
			\begin{tabular}{wc{1.5cm}wc{1.8cm}wc{2cm}wc{3cm}}
				\multicolumn{4}{c}{$L=8$} \\
				\hline
				\hline
				$m$ &  $w_d$  & $E_0(m)$ &  $E_0(m)-E_0(\infty)$ \\
				\hline
				2000   &  1.286e-05  &  -31.76311291   &   0.26322934 \\
				4000   &   8.420e-06  &  -31.87302879   &   0.15331346 \\	
				6000   &  5.681e-06  &  -31.91448613   &   0.11185612 \\
				8000   &   4.550e-06  &  -31.93686464   &   0.08947761 \\
				10000   &  4.165e-06  &  -31.95093410   &   0.07540815 \\
				12000   &  3.815e-06  &  -31.96253340   &   0.06380885 \\
				\hline
				$\infty$ &             &   -32.02634225  &            \\
				\hline
				\hline
			\end{tabular}
		\end{center}
	\end{minipage}
	\hfill
	\begin{minipage}[t]{.48\textwidth}
		\begin{center}
			\begin{tabular}{wc{1.5cm}wc{1.8cm}wc{2cm}wc{3cm}}
				\multicolumn{4}{c}{} \\
				\multicolumn{4}{c}{$L=10$} \\
				\hline
				\hline
				$m$ &  $w_d$  & $E_0(m)$ &  $E_0(m)-E_0(\infty)$ \\
				\hline
				2000   &  2.119e-05  &  -49.50077175  &  1.02132882 \\
				4000   &  1.379e-05  &  -49.84727132  &  0.67482925 \\
				6000   &  1.198e-05  &  -49.99628942  &  0.52581115 \\
				8000   &  9.848e-06  &  -50.09485526  &  0.42724531 \\
				10000   &  7.774e-06  &  -50.15642914  &  0.36567143 \\
				12000   &  5.854e-06  &  -50.19615496  &  0.32594561 \\
				\hline
				$\infty$ &             &   -50.52210057  &            \\
				\hline
				\hline
			\end{tabular}
		\end{center}
	\end{minipage}
	\hfill
	\begin{minipage}[t]{.48\textwidth}
		\begin{center}
			\begin{tabular}{wc{1.5cm}wc{1.8cm}wc{2cm}wc{3cm}}
				\multicolumn{4}{c}{} \\
				\multicolumn{4}{c}{$L=12$} \\
				\hline
				\hline
				$m$ &  $w_d$  & $E_0(m)$ &  $E_0(m)-E_0(\infty)$ \\
				\hline
				2000  &  -   &    -  &   - \\
				4000  &   1.840e-05  &  -71.40452905   &  1.73164686 \\
				6000  &   1.486e-05  &  -71.71734724   &  1.41882867 \\
				8000  &   1.365e-05  &  -71.91712273   &  1.21905318 \\
				10000  &   1.224e-05  &  -72.06942431   &  1.06675160 \\
				12000  &   9.293e-06  &  -72.19219444   &  0.94398147 \\
				\hline
				$\infty$ &             &   -73.13617591  &            \\
				\hline
				\hline
			\end{tabular}
		\end{center}
	\end{minipage}
	\label{SBCdata}
\end{table}

We here check the accuracy of DMRG calculations under SBC (periodic chain).
In Table~\ref{SBCdata} the Discarded weight $w_d$ and the ground-state energy
$E_0(m)$ of the half-filled square-lattice Hubbard model at $U/t=8$ are
shown as a function of the number of density-matrix eigenstates kept in
the renormalization $m$ for various system sizes $L \times L$. For all
cases the ground-state energy at $m\to\infty$ ($w_d\to0$) can be accurately
obtained by a linear fit of $E_0(m)$ vs. $w_d$. We also find that the
discarded weight is roughly proportional to $L$, i.e., $w_d \propto L$. Nevertheless,
the discarded weight is still order of $\sim10^{-5}$ even for $12 \times 12$.
This may fall into the category of an accurate DMRG calculation for 2D fermionic system.

In Table~\ref{DMRGdata} the discarded weight in the DMRG calculation for
the square-lattice Heisenberg model is compared between various boundary
conditions. The discarded weight is almost comparable for OBC and cylinder,
and somewhat larger for the SBC-projected periodic chain. Surprisingly,
it is much smaller when the SBC-projected open chain is studied. Thus,
from the viewpoint of discarded weight, OBC, SBC, and cylinder can be equally
good options in DMRG calculation for 2D systems

\newpage

\begin{table}[tbh]
	\caption{Discarded weight $w_d$ and the ground-state energy $E_0(m)$
		of the square-lattice Heisenberg model as a function of the number
		of density-matrix eigenstates kept in the renormalization $m$ for
		system sizes ($6 \times 6$ ($L=6$) and $8 \times 8$ ($L=8$) with
		various boundary conditions. The ground-state energy at $m\to\infty$
		was obtained by a linear fit of $E_0(m)$ vs. $w_d$.}
	\begin{minipage}[t]{.90\textwidth}
		\begin{center}
			\begin{tabular}{wc{1.8cm}|wc{1.8cm}wc{2cm}wc{2.5cm}|wc{1.8cm}wc{2cm}wc{2.5cm}}
				\multicolumn{7}{c}{OBC} \\
				\hline
				\hline
				&\multicolumn{3}{c}{$L=6$}&\multicolumn{3}{c}{$L=8$} \\
				$m$&$w_d$&$E_0(m)$&$E_0(m)-E_0(\infty)$&$w_d$&$E_0(m)$&$E_0(m)-E_0(\infty)$ \\
				\hline
				1000&1.100e-07&-21.72673267&0.00005340&8.166e-06&-39.60759545&0.01078493\\
				2000&6.466e-09&-21.72678301&0.00000307&1.587e-06&-39.61627493&0.00210544\\
				3000&9.883e-10&-21.72678556&0.00000051&5.714e-07&-39.61762743&0.00075294\\
				4000&2.279e-10&-21.72678592&0.00000015&2.636e-07&-39.61803809&0.00034228\\
				\hline
				&         &-21.72678607&          &         &-39.61838038&          \\
				\hline
				\hline
			\end{tabular}
		\end{center}
	\end{minipage}
	
	\vspace*{5.0mm}
	
	\begin{minipage}[t]{.90\textwidth}
		\begin{center}
			\begin{tabular}{wc{1.8cm}|wc{1.8cm}wc{2cm}wc{2.5cm}|wc{1.8cm}wc{2cm}wc{2.5cm}}
				\multicolumn{7}{c}{PBC} \\
				\hline
				\hline
				&\multicolumn{3}{c}{$L=6$}&\multicolumn{3}{c}{$L=8$} \\
				$m$&$w_d$&$E_0(m)$&$E_0(m)-E_0(\infty)$&$w_d$&$E_0(m)$&$E_0(m)-E_0(\infty)$ \\
				\hline
				1000&2.981e-05&-24.40302953&0.03665799&-&-&-\\
				2000&9.188e-06&-24.42949472&0.01019281&-&-&-\\
				3000&3.938e-06&-24.43541285&0.00427468&-&-&-\\
				4000&1.982e-06&-24.43743236&0.00225516&-&-&-\\
				\hline
				&         &-24.43968752&          & & &\\
				\hline
				\hline
			\end{tabular}
		\end{center}
	\end{minipage}
	
	\vspace*{5.0mm}
	
	\begin{minipage}[t]{.90\textwidth}
		\begin{center}
			\begin{tabular}{wc{1.8cm}|wc{1.8cm}wc{2cm}wc{2.5cm}|wc{1.8cm}wc{2cm}wc{2.5cm}}
				\multicolumn{7}{c}{SBC (periodic chain)} \\
				\hline
				\hline
				&\multicolumn{3}{c}{$L=6$}&\multicolumn{3}{c}{$L=8$} \\
				$m$&$w_d$&$E_0(m)$&$E_0(m)-E_0(\infty)$&$w_d$&$E_0(m)$&$E_0(m)-E_0(\infty)$ \\
				\hline
				1000&9.130e-06&-24.45300594&0.00801971&9.025e-05&-42.79118365&0.32899812\\
				2000&1.661e-06&-24.45965335&0.00137230&4.565e-05&-42.96307806&0.15710371\\
				3000&5.365e-07&-24.46058729&0.00043835&2.898e-05&-43.02157949&0.09860228\\
				4000&2.260e-07&-24.46083548&0.00019017&1.973e-05&-43.05183014&0.06835163\\
				\hline
				$\infty$&         &-24.46102565&          &         &-43.12018177&\\
				\hline
				\hline
			\end{tabular}
		\end{center}
	\end{minipage}
	
	\vspace*{5.0mm}
	
	\begin{minipage}[t]{.90\textwidth}
		\begin{center}
			\begin{tabular}{wc{1.8cm}|wc{1.8cm}wc{2cm}wc{2.5cm}|wc{1.8cm}wc{2cm}wc{2.5cm}}
				\multicolumn{7}{c}{SBC (open chain)} \\
				\hline
				\hline
				&\multicolumn{3}{c}{$L=6$}&\multicolumn{3}{c}{$L=8$} \\
				$m$&$w_d$&$E_0(m)$&$E_0(m)-E_0(\infty)$&$w_d$&$E_0(m)$&$E_0(m)-E_0(\infty)$ \\
				\hline
				1000&1.037e-09&-23.07464567&0.00000050&9.819e-07&-41.31375095&0.00116164\\
				2000&2.447e-11&-23.07464616&0.00000001&1.161e-07&-41.31476448&0.00014811\\
				3000&2.351e-12&-23.07464617&$<10^{-8}$&2.987e-08&-41.31487486&0.00003773\\
				4000&3.837e-13&-23.07464618&$<10^{-8}$&1.058e-08&-41.31489881&0.00001379\\
				\hline
				$\infty$&         &-23.07464618&          &         &-41.31491259&\\
				\hline
				\hline
			\end{tabular}
		\end{center}
	\end{minipage}
	
	\vspace*{5.0mm}
	
	\begin{minipage}[t]{.90\textwidth}
		\begin{center}
			\begin{tabular}{wc{1.8cm}|wc{1.8cm}wc{2cm}wc{2.5cm}|wc{1.8cm}wc{2cm}wc{2.5cm}}
				\multicolumn{7}{c}{cylinder} \\
				\hline
				\hline
				&\multicolumn{3}{c}{$L=6$}&\multicolumn{3}{c}{$L=8$} \\
				$m$&$w_d$&$E_0(m)$&$E_0(m)-E_0(\infty)$&$w_d$&$E_0(m)$&$E_0(m)-E_0(\infty)$ \\
				\hline
				1000&3.349e-08&-23.09009340&0.00001678&4.299e-06&-41.34358356&0.00569929\\
				2000&1.512e-09&-23.09010942&0.00000076&7.206e-07&-41.34831869&0.00096415\\
				3000&2.012e-10&-23.09011008&0.00000010&2.330e-07&-41.34898086&0.00030198\\
				4000&4.556e-11&-23.09011016&0.00000002&9.196e-08&-41.34916154&0.00012131\\
				\hline
				$\infty$&         &-23.09011018&          &         &-41.34928284&\\
				\hline
				\hline
			\end{tabular}
		\end{center}
	\end{minipage}
	\label{DMRGdata}
\end{table}

\newpage

\section{VIII. Advantages of spiral boundary conditions}

\begin{figure}[h]
	\includegraphics[width=0.9\linewidth]{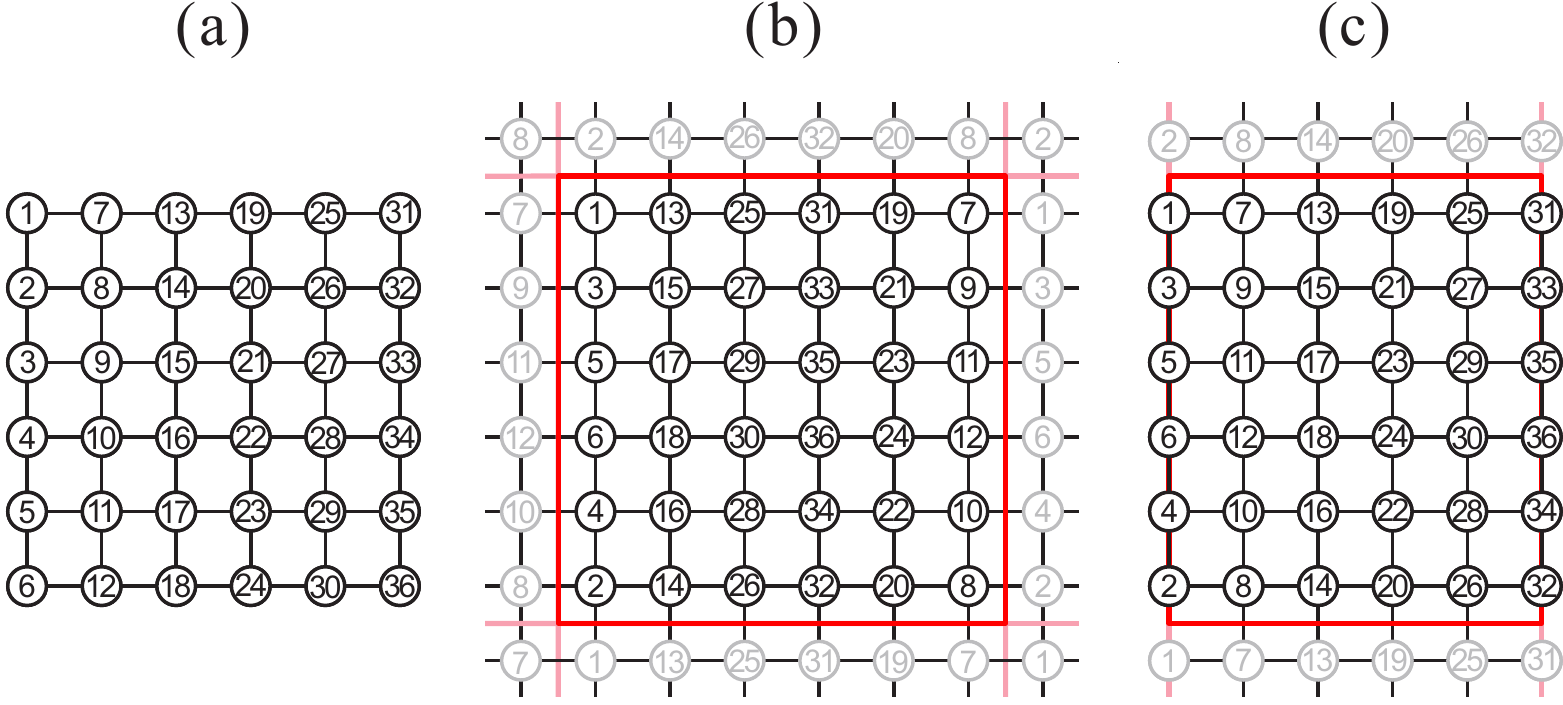}
	\caption{2D square-lattice $6\times6$ cluster with (a) OBC, (b) PBC,
		and (c) cylinder. In each cluster an optimal order of sites for
		DMRG simulation is denoted by numbers.
	}
	\label{fig:lattice6x6_SM}
\end{figure}

In DMRG simulations we have to choose a proper boundary condition for
considered system and quantities. The applicability of DMRG to a lattice
system can be roughly judged by the distance between two sites connected
by the longest-range term in the system. The distance is denoted by $d$ below.
In general, the DMRG calculation is difficult if $d$ is larger than $\sim10$.
Let us then estimate the values of $d$ for various boundary conditions. In
Fig.~\ref{fig:lattice6x6_SM} an optimal order of sites to give the shortest
$d$ for a 2D square-lattice cluster is illustrated for each boundary condition. 
When we use OBC or cylinder, $d$ is $L$ for a $L \times L$ cluster. This is comparable to that for SBC ($d=L-1$ or $d=L$). However, if we use PBC, $d$
jumps to $2L$. This is definitely ill-conditioned for DMRG treatments.
For example, the system size is limited up to $L\sim6$ in the Heisenberg model.
Therefore, PBC may be firstly excluded from the choice, although it is
naively thought to be a most versatile boundary condition in numerical
calculations.

In fact, either OBC or cylinder is typically employed in the previous DMRG
simulations. These boundary conditions are technically rather convenient
for DMRG simulations because the value of $d$ can be reduced down to $L$
with the optimal order of sites. Nevertheless, they could often lead to some
sorts of problems in the physical properties. The followings are examples of
issues: With OBC, the bulk state can be significantly disturbed by Friedel
oscillations, and also the properties of bulk and edge states can be completely
different due to the missing outside bonds especially in doped systems.
These problems could be managed by controlling open edges for a specific
states, and however, it is not always successful. With cylinder, 
a kind of spatial anisotropy is introduced in spite of the original isotropic
2D square-lattice cluster. A most problematic point is that short loops of
bonds are formed along the circumference direction. This could results in
an unnatural periodicity of the wave function as well as an unexpected
plaquette constraint of particles or spins. Accordingly, as shown in the
previous section of Supplemental Material, the density-density correlation
functions can be largely disturbed by cylindrical boundary.

With SBC, Friedel oscillations can be avoidable and no artificial short loops
of bonds are created. Also, $d$ is only $L$ or $L-1$. Furthermore,
a SBC-projected chain with open boundary is used, the accuracy of DMRG
calculation is even better than that with OBC and cylinder as shown in the
previous section of Supplemental Material. Thus, SBC enables us to perform
accurate DMRG calculations with skirting some sorts of finite-size effects
seen in OBC and cylindrical clusters. However, we expect that SBC does not
always provide better results. For example, OBC may be more appropriate to
study a system exhibiting valence-bond-solid state such as plaquette/bond
order; and, a cylindrical cluster may be more appropriate to study a system
exhibiting magnetic/charge/bond ordered state with spatial rotation symmetry
breaking.

We thus think that SBC can be a main option of the boundary conditions in
DMRG calculations for the above technical and physical reasons.